# Learning to deform the matched filter

Paul Anthony Haigh

**Analytical signal-processing blocks are interpretable and reliable, but their optimality depends on assumptions that practical hardware and channels violate. Learned replacements can adapt, but often discard the structure that makes the original solution understandable. Here we introduce deformable matched filtering, a platform in which learning steers a bounded deformation of an explicit matched filter instead of replacing the waveform-processing path. In a hardware-in-the-loop optical wireless link, blind state descriptors drive causal, pilotless updates while payload samples, transmitted references, and condition labels remain outside the controller. Receiver deformation generalises across held-out signalling and channel conditions and remains ahead of a span-matched fractionally spaced equaliser after stationary convergence in most tested regimes. A transferable transmitter deformation reduces error-vector magnitude in all 144 held-out evaluations, whereas its additional value after receiver adaptation emerges principally under severe combined distortion. KAN, MLP, and linear controllers provide different condition-dependent advantages within the same filter structure. During 24 hours of uninterrupted changing-condition operation, bounded deformable receivers recover their original operating regime after severe intervening distortion while the persistent conventional equaliser accumulates destructive state. These results establish deformable analytical filters as a reusable middle ground between fixed theory and end-to-end learned signal processing.**

Matched filtering is a foundational analytical solution in communication theory: for a known waveform in additive white Gaussian noise, it maximises output signal-to-noise ratio (1). Practical links violate those assumptions through bandwidth limitation, nonlinear transfer, timing uncertainty, memory, component drift, and mismatch between nominal and realised responses. The conventional filter remains physically interpretable, but a fixed implementation cannot follow these departures.

Adaptive equalisers address mismatch by estimating a new filter, while learned physical-layer systems can replace larger portions of the transmitter, receiver, or both (2-6). A complementary line of research embeds learning in known algorithms or physical models through model-based deep learning and algorithm unrolling (7, 8), learned digital backpropagation (9, 10), and globally trainable conventional DSP blocks (11). Recent optical-wireless experiments have likewise demonstrated neural post-equalisation and hardware-interactive transmitter pre-equalisation (12, 13). These approaches can be powerful, but they assign adaptation to a new signal-processing object. That choice can obscure which part of the analytical solution remains useful, complicate causal operation, and entangle adaptation with payload recovery.

Our earlier work introduced a supervised deformable matched filter (DMF) for carrier-less amplitude-and-phase signalling under bandwidth limitation (14). That study used known transmitted symbols to optimise EVM in a restricted channel setting. The present work asks a substantially broader question: whether the matched filter can remain a continuously adaptive platform during ordinary transmission, without pilots, while generalising across signalling families, nonlinear and combined impairments, transmitter-receiver allocation, and persistent hardware operation.

We retain the analytical solution as a structural prior and learn only how it should deform. The zero-deformation state is the conventional matched filter, every adaptive state remains a measurable departure from it, and different controllers can be compared without changing the waveform-processing path. This formulation turns model choice into a controller-design question over a comparison between different receivers.

Contrary to most communication systems, learning does not operate on the communication waveform itself. Each received block is compressed into a frozen set of blind, physically meaningful state descriptors. A Kolmogorov-Arnold network (KAN), multi-layer perceptron (MLP), or linear controller maps those descriptors to a low-dimensional filter deformation that is applied only to subsequent data. Payload samples and transmitted reference symbols are never controller inputs, preserving a strict causal boundary between observation, update, and action while extending the training-free principle of classical blind equalisation (15).

This architecture asks a general scientific question: can a known analytical solution learn continuously while preserving its structure, physical interpretation, and role in the signal-processing chain? We test that question using receiver deformation, a similarly constrained transmitter pulse, and their two-sided composition. Because transmitter shaping acts before information is lost or distorted and receiver adaptation acts after the realised channel, their relative value reveals where compensation is physically most effective.

We use a hardware-in-the-loop (HIL) optical wireless link as a controlled experimental platform, but the transmission technology is agnostic and not a requirement of the method. The controller receives no optical-channel identity or impairment label and is completely blind to the specifics of the test setup. We test PAM-2 and PR4 signals, held-out pulse shapes, bandwidth-only and nonlinear distortions, combined impairments, abrupt switching, randomised stationary dwells, and a 24-hour changing-condition run. A fairly designed decision-directed fractionally spaced equaliser (FSE) provides a conventional adaptive-filter benchmark.

The experiments reveal a platform with advantages and trade-offs, rather than a universal winning learner. Receiver deformation produces broad gains while transmitter deformation is largely captured by a transferable static pulse. Their complementarity emerges under severe combined distortion, and the preferred transmitter-receiver allocation depends on controller architecture. Linear control, KAN, and MLP each occupy useful operating niches. Stationary-dwell controls and the full-day return experiment show that the result is not an artifact of per-condition retraining, rapid switching, or an unconverged conventional equaliser.

## RESULTS

### Matched filters adapt without being replaced

We first tested whether a matched-filter receiver could adapt continuously during communication while retaining the matched filter as its explicit signal-processing structure. We represented the receive filter at adaptation block *k* as:

$$\mathbf{h}_{\mathbf{Rx},k} = \mathbf{h}_{\mathbf{0}} + \Delta\mathbf{h}_{\mathbf{Rx},k} \tag{1}$$

where $\mathbf{h}_{\mathbf{0}}$ denotes the nominal matched-filter impulse response and $\Delta\mathbf{h}_{\mathbf{Rx},k}$ denotes a constrained learned deformation. The zero-deformation state therefore corresponds exactly to the conventional matched filter rather than to an independently learned receiver.

We imposed three restrictions on adaptation. First, each logical link began from the conventional matched-filter solution. Second, the learning model never received the communication waveform directly and never performed symbol recovery. Instead, we extracted a compact set of blind receiver-state features from each observed signal block and supplied only these descriptors to the controller, which used them to modify the parameters of the deformable matched filter. The controller received no pilot sequence, transmitted payload reference, impairment identity, or modulation-specific control signal and therefore adapted only from signal-state information available at the receiver. Third, observations collected during block could influence only subsequent blocks:

$$\mathcal{O}_k \longrightarrow \Delta\mathbf{h}_{\mathbf{Rx},k} + 1 \tag{2}$$

so that information extracted from one block could not alter the filter used to recover that same block. Receiver state persisted between successive hardware-in-the-loop rounds, allowing adaptation to accumulate as transmission continued.

We compared three controllers within the same receiver structure: a KAN, an MLP, and a linear model. Each received the same frozen 26-element blind state vector and controlled the same 64-mode deformation family around a 301-tap matched filter. The controllers therefore differed only in how blind state was mapped to allowed filter displacement. As conventional baselines, we used the fixed matched filter and a decision-directed FSE whose temporal span was designed to match the deformable receiver. A direct neural waveform equaliser would test a different architectural hypothesis by replacing the explicit filtering path; the FSE instead provides a like-for-like adaptive-filter benchmark. The KAN configurations use learnable univariate edge functions rather than fixed node activations (16). We next tested whether the prespecified deformable receivers generalised beyond their development conditions. Controller architecture, blind descriptors, deformation space, learning rules, and stability controls were fixed before qualification, while controller and filter state continued to update causally. We evaluated PAM-2 and PR4 signalling at unseen root-raised-cosine roll-offs of 0.15, 0.25, and 0.35 under clean, bandwidth-limited, nonlinear, and combined bandwidth-nonlinear regimes (Physical channel impairments in Methods; Fig. 1).

For PR4 signalling, all three deformable receivers improved on the fixed matched filter across the held-out roll-offs, but the preferred controller depended on the physical distortion (Fig. 1B). KAN was particularly strong under clean and nonlinear conditions, whereas linear control often led under bandwidth-limited and combined conditions; MLP closely followed KAN across much of the test space. The deformable structure therefore transferred even when the pulse shape and impairment differed from receiver development, while the best mapping from blind state to deformation remained regime dependent.

The same qualitative behaviour extended to PAM-2 (Fig. 1C). Receiver deformation again reduced EVM relative to the fixed matched filter and generally remained ahead of the span-matched FSE, while the ordering of KAN, MLP, and linear control changed with condition. A condition-resolved summary relative to the FSE is provided in Supplementary Fig. S4. The shared advantage therefore lies in constraining adaptation to an explicit family of matched-filter deformations, not in one controller being universally superior.

An independently acquired partial physical repeat reproduced the clean-condition controller trajectories (Supplementary Fig. S3). The small, synchronised excursions were also reproducible and coincided with activation of the prespecified late-trust constraint, which projects accumulated displacement onto its calibrated bound before subsequent causal readaptation (Supplementary Fig. S9). These diagnostics support a structured receiver-state interpretation without requiring

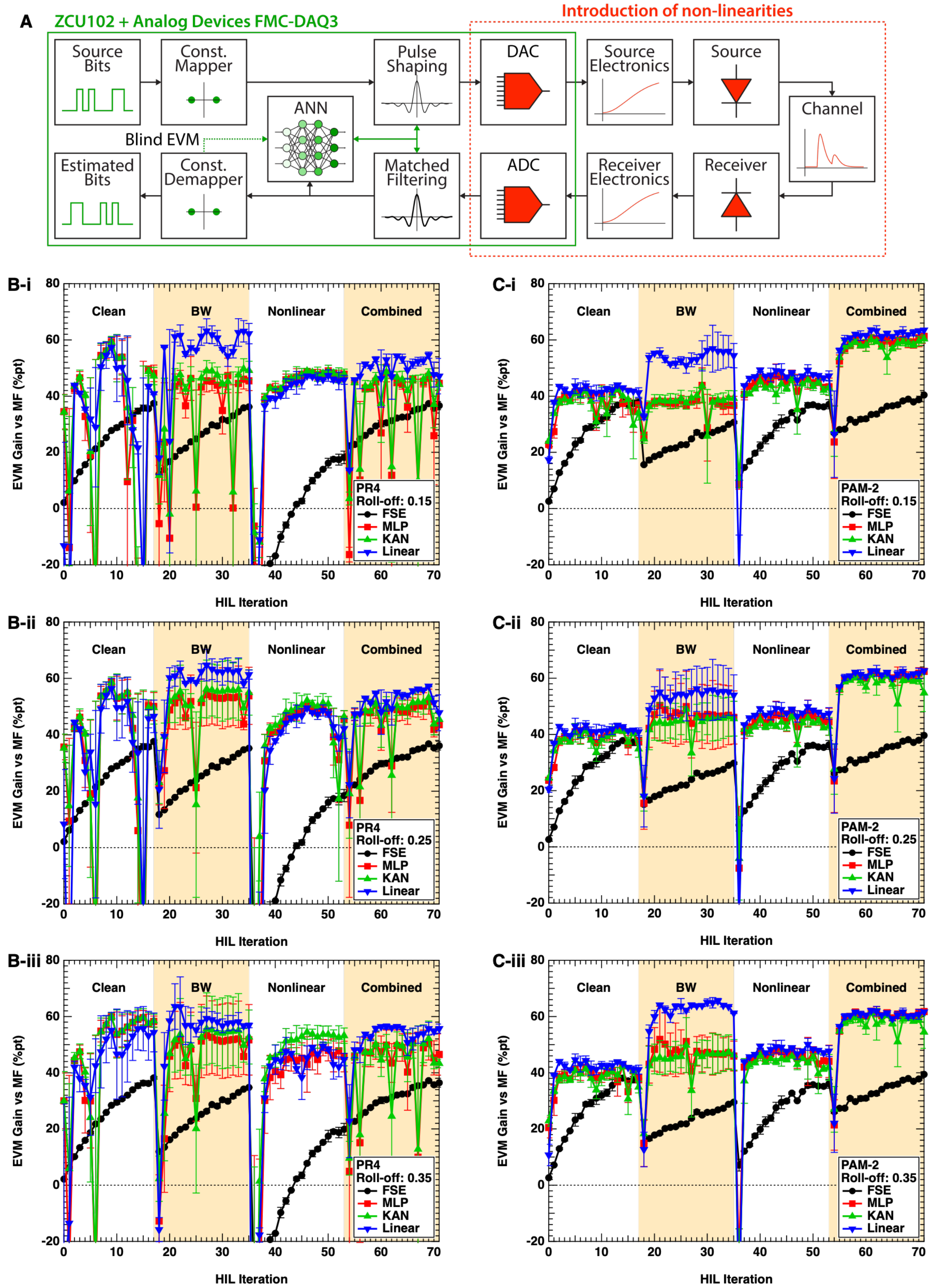

**Fig. 1. Causal deformation of an explicit matched-filter receiver generalises across held-out waveforms and impairments.** (A) Hardware-in-the-loop optical link and causal adaptive receiver. The controller shown schematically as an ANN represents the KAN, MLP, or linear mapping from blind receiver-state descriptors to matched-filter deformation; payload references and condition labels are unavailable to online adaptation. (B-i to B-iii) EVM improvement in percentage points relative to the fixed matched filter for PR4 signalling at held-out RRC roll-offs of 0.15, 0.25, and 0.35. (C-i to C-iii) Corresponding PAM-2 qualification. Each sequence traverses clean, bandwidth-limited, nonlinear, and combined impairment epochs without a condition-triggered reset. DMF points show means across four persistent controller seeds on matched physical captures; error bars show seed standard deviations. Positive values indicate lower EVM than the fixed matched filter.

the main qualification figure to carry adaptation-state detail.

Together, these experiments show that the matched filter can remain the explicit receiver while learning modifies only its operating point. The learned models do not replace symbol recovery; they determine how the analytical filter should deform from one causal block to the next.

### Transmitter deformation is largely static

Having established that the receiver could adapt causally and persistently while retaining the matched-filter structure, we next asked whether the same principle could be applied at the transmitter. Rather than learning an arbitrary transmitted waveform, we constrained the transmitted pulse to a low-dimensional deformation of the nominal root-raised-cosine pulse:

$$\mathbf{h}_{\mathrm{Tx}} = \mathbf{h}_{\mathrm{RRC}} + \mathbf{B}_{\mathrm{Tx}}\mathbf{c} \qquad (3)$$

where $\mathbf{B}_{\mathrm{Tx}}$ contained eight smooth basis functions and $\mathbf{c}$ defined their amplitudes. This retained the conventional pulse as an explicit reference while allowing its shape to move within a bounded family of physically realisable responses (Fig. 2A). The receiver used for this experiment was fixed to the canonical matched-filter path so that any change in performance could be attributed to the transmitted pulse rather than to simultaneous receiver adaptation.

We first optimised a single transmitter deformation across PAM-2 and PR4 signalling in a prescribed five-condition development cycle (Physical channel impairments in Methods). The resulting pulse was frozen and evaluated without further optimisation at previously unseen root-raised-cosine roll-offs of 0.15, 0.25, and 0.35 under the same four held-out impairment regimes used for receiver qualification (Fig. 2). Relative to the conventional transmitter, the frozen pulse reduced mean EVM by 16.97, 16.723 and 16.37%, respectively. The gain remained stable as the held-out experiment traversed all four impairment classes (Fig. 2B).

Across all 144 paired held-out evaluations, the deformed transmitter produced lower EVM than the conventional pulse (Fig. 2C and Supplementary Fig. S5). The close agreement across roll-offs indicates that the learned pulse did not compensate only one nominal waveform or one impairment, but captured a broadly transferable transmitter operating point.

We separately tested whether the transmitter needed to continue adapting during operation. A small residual deformation was controlled from six delayed receiver-side blind memory descriptors; feedback from one HIL round could affect only later transmissions. Continued adaptation added mean EVM reductions of 12.5% for KAN, 14.2% for MLP, and 15.7% for linear control beyond the frozen pulse (Supplementary Fig. S5C). The contrast with the receiver was marked: transmitter deformation was valuable, but nearly all of its benefit was captured statically.

These results expose two adaptation timescales within the same platform. The transmitter can establish a broadly favourable waveform before propagation, whereas the receiver remains directly exposed to changing channel and hardware distortions and benefits from persistent causal tracking. Subsequent two-sided experiments therefore used the frozen transmitter pulse while allowing the receiver to remain adaptive.

### Two-sided deformable-filter platform reveals condition-dependent complementarity

Having established the receiver and transmitter implementations separately, we next used the same deformable matched-filter platform to ask how compensation should be divided between the two ends of a link. The platform preserves an explicit analytical signal-processing object at both ends: a transmitted pulse deformed around its nominal root-raised-cosine form and a receive filter deformed around the corresponding matched filter. The receiver-side deformation can be steered by KAN, MLP, or linear controllers without changing the waveform-processing path. We first compared the four factorial configurations formed by conventional or deformed transmission and fixed or adaptive reception. Before evaluation, we fixed the transmitter pulse, receiver architecture, state descriptors, deformation space, learning rules, and stability controls; receiver weights and filter state nevertheless continued to update causally during transmission. The transmitter and receiver were not jointly re-optimised for these tests.

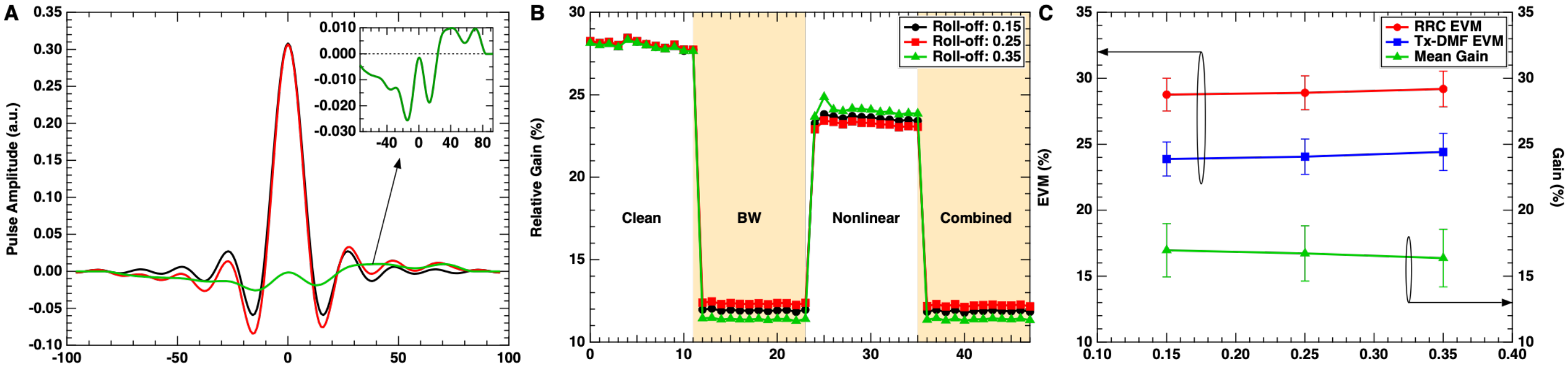

**Fig. 2. A transferable static pulse captures most transmitter-side benefit.** (A) Nominal RRC pulse (black), physically evaluated frozen Tx-DMF pulse (red), and their signed deformation (green; inset expands the deformation). (B) Paired relative EVM reduction of the same frozen pulse during held-out clean, bandwidth-limited, nonlinear, and combined conditions at RRC roll-offs of 0.15, 0.25, and 0.35. (C) Mean conventional-transmitter EVM, Tx-DMF EVM, and relative gain for each held-out roll-off (48 paired evaluations per roll-off). EVM error bars show standard errors across evaluations; gain error bars show 95% t-confidence intervals. Detailed paired results and the residual online-transmitter experiment are shown in Supplementary Fig. S5.

The initial factorial experiment established a useful boundary condition. With the fixed matched-filter receiver, Tx-DMF reduced mean EVM from 29.68% to 24.98%, a 15.8% relative reduction observed in all 20 HIL rounds. Once receiver deformation was active, however, the same transmitter pulse produced no material further reduction: mean EVM changed from 20.61% to 20.56% for KAN, from 20.80% to 20.95% for MLP, and from 17.65% to 18.14% for linear control. Under these moderate conditions, transmitter and receiver deformation were therefore largely redundant rather than additive (Supplementary Fig. S6).

We next tested whether this balance changed as the channel became more restrictive. The same prespecified transmitter pulse and receiver deformation family were evaluated across progressively stronger bandwidth limitation, nonlinear distortion, and combined distortion without changing features, hyperparameters, constraints, or adaptation policy (Fig. 3A). MLP gave the largest incremental transmitter benefit through the bandwidth-only sweep, linear control was strongest across much of the nonlinearity-only sweep and at the intermediate combined condition, and KAN was strongest at the hardest combined condition. At normalised bandwidth 0.475 and nonlinear strength 0.90, adding Tx-DMF reduced EVM by 4.00% for KAN, 3.00% for MLP, and 1.28% for linear control relative to the corresponding receiver-only link.

The controller-by-condition map shows why no model in isolation should be treated as the platform itself (Fig. 3A). Small positive and negative changes occurred under isolated mild impairments, but the clearest common progression occurred under combined distortion. Averaged descriptively across controllers, incremental transmitter benefit increased from 0.26% at the mildest combined condition to 2.01% at the intermediate condition and 2.76% at the most severe condition (Fig. 3B). Transmitter shaping therefore becomes useful when propagation removes structure that receiver-side processing alone cannot fully recover.

This behaviour reveals an asymmetric form of transmitter-receiver complementarity within a common platform. When distortion is modest, the adaptive receiver can compensate most of the recoverable mismatch after propagation. As bandwidth restriction and nonlinearity become more severe, shaping the waveform before transmission becomes increasingly valuable because some channel-induced loss is difficult to undo retrospectively. This observation motivated the next experiment, in which we varied the allowed transmitter and receiver deformation budgets to determine how compensation should be distributed between the two sides of the link.

## Compensation has no universal allocation

The preceding experiments showed that transmitter deformation becomes more useful as channel distortion increases, but not how the available compensation should be divided between the two ends. We therefore varied the allowed transmitter and receiver deformation budgets while leaving the transmitter direction, blind state descriptors, controller architectures, learning rules, and stability controls unchanged (Fig. 4).

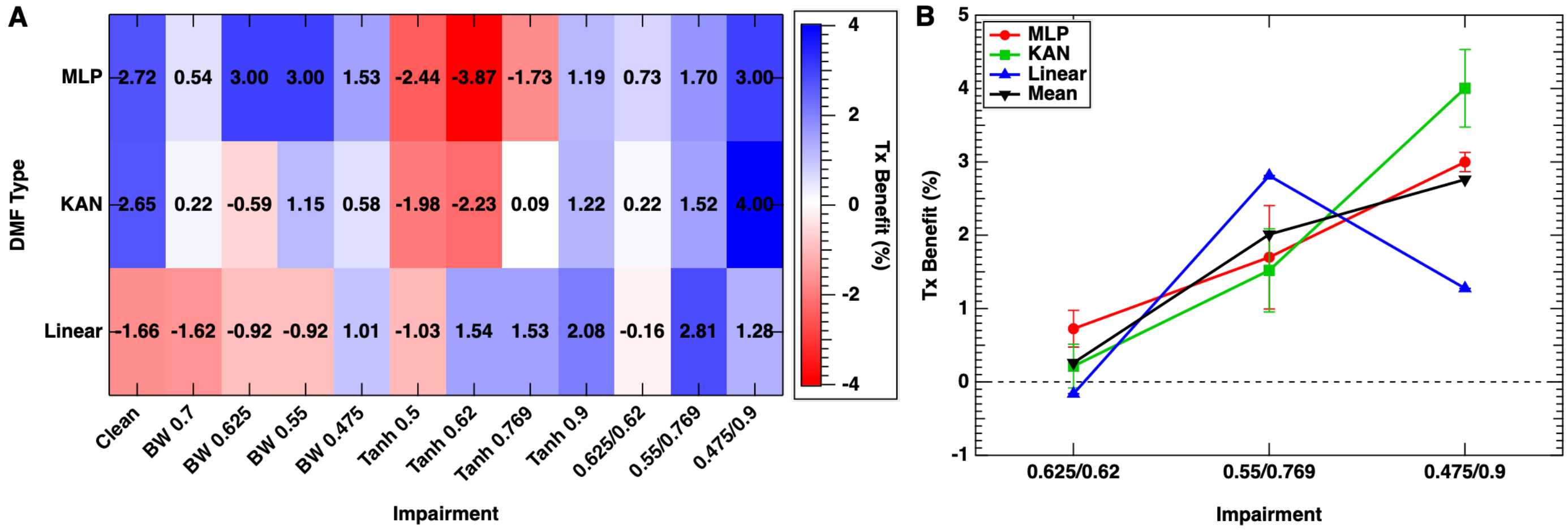

**Fig. 3. Controller-specific transmitter-receiver complementarity emerges with distortion severity.** (A) Incremental relative EVM reduction produced by adding the fixed Tx-DMF pulse to the corresponding adaptive Rx-DMF link. Columns group clean, bandwidth-only, nonlinearity-only, and combined impairments in severity order; positive values indicate lower EVM with transmitter deformation. KAN, MLP, and linear controllers use the same blind descriptors and explicit receiver-filter family, so differing patterns indicate controller specialisation. (B) Controller-resolved benefit across the three combined bandwidth/nonlinearity conditions. Error bars show standard deviations across four persistent controller seeds; the black line is the descriptive mean across controller architectures.

We evaluated six transmitter scales from 0 to 1.25 and four receiver scales from 0.50 to 1.25 under three combined bandwidth-nonlinear conditions. The point (1.00,1.00) reproduced the independently determined settings, whereas transmitter scale 0 provided receiver-only references. Each transmitter setting was a separate physical acquisition, and its capture was reused across receiver scales so that comparisons along the receiver axis were paired to the same physical waveform.

No allocation was best for every controller (Fig. 4A-C). KAN and MLP reached their lowest sampled mean EVM at transmitter scale 0.75 and receiver scale 1.25, improving on the common (1.00,1.00) reference by 4.23% and 3.10%, respectively. Because these minima lie at the sampled receiver boundary, they are best mapped points rather than evidence of an interior optimum. Linear control behaved differently: its best mapped point occurred at transmitter scale 1.00 and receiver scale 0.50, 10.17% below the common reference.

Transmitter deformation near three quarters of the independently determined level was favoured for KAN and MLP, but the preferred receiver budget depended strongly on controller architecture. Nonlinear controllers benefited from greater receiver deformation within the tested range, whereas linear control favoured substantially less. The allocation problem therefore cannot be reduced to maximising adaptation at both ends or placing all compensation at the receiver.

These results also clarify the role of the independently optimised transmitter and receiver settings used elsewhere in the study. The $(1,1)$ operating point provides a common, architecture-independent reference, but it is not necessarily the best joint allocation once the two sides interact. At the same time, the boundary-limited optima for KAN and MLP caution against interpreting the mapped surface as a globally optimised solution. The experiment instead demonstrates that transmitter and receiver deformation form a coupled compensation problem whose preferred balance depends on the structure of the adaptive controller.

Because architecture-specific retuning would confound direct system comparison, subsequent persistent-link experiments returned to the common (1.00,1.00) operating point. This preserved a neutral reference while testing whether the platform remained useful over persistent physical interactions.

We also performed an intentionally adversarial 180-round sequence in which the impairment changed nearly every round while state persisted. It verified bounded state carry-over but neither allowed meaningful convergence nor approximated realistic channel evolution, and is therefore reported only as a supplementary stress control (Supplementary Fig. S7).

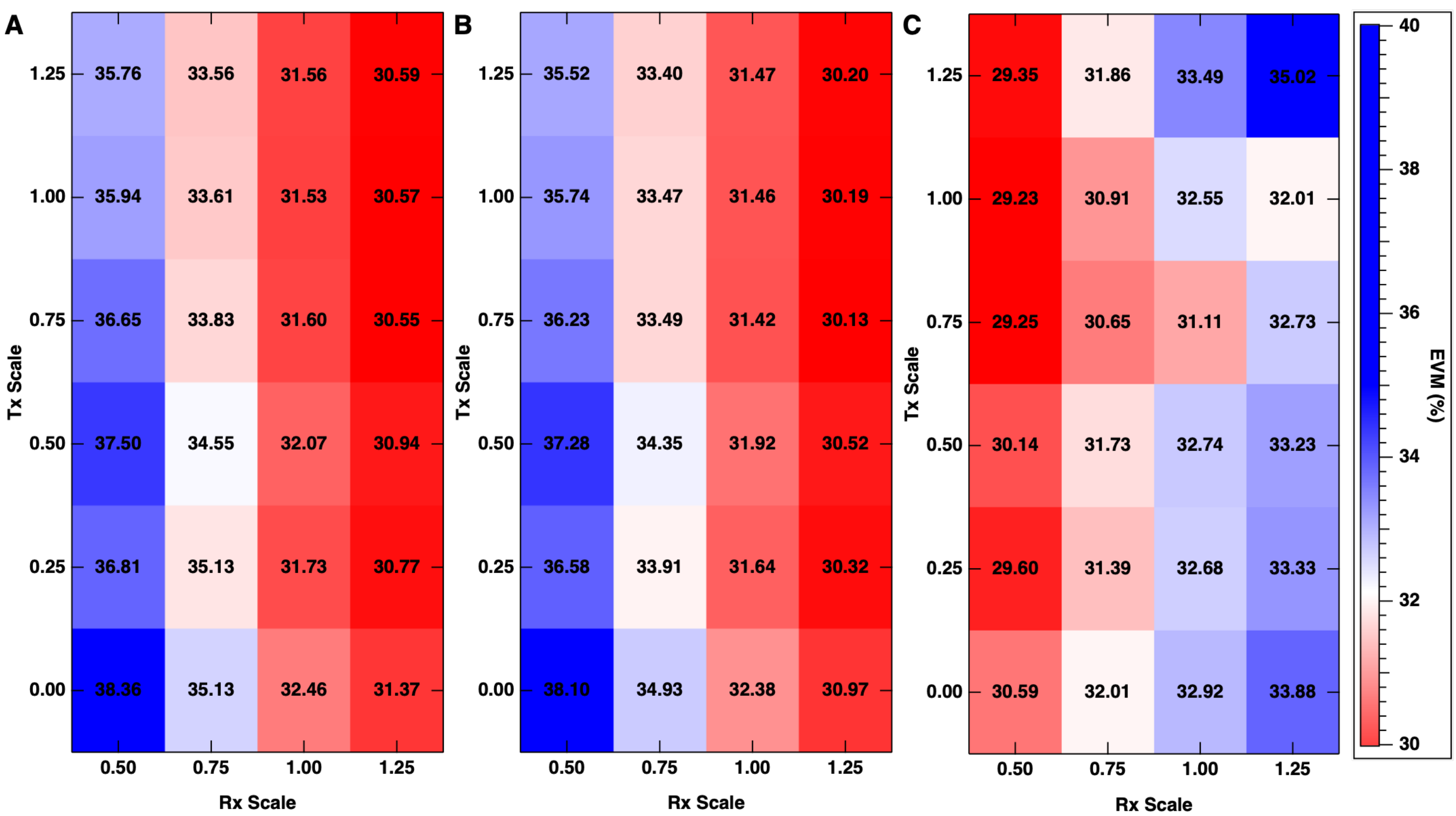

**Fig. 4. Transmitter-receiver compensation has no universal allocation.** Mean EVM surfaces for (A) KAN, (B) MLP, and (C) linear control as the frozen transmitter-deformation scale and allowed receiver-deformation scale are varied. Each cell first averages three physical repeats within each of three combined-impairment conditions and controller seed, then averages the resulting 12 condition-seed means. All receiver features, controller architecture, update rules, and stability controls remain unchanged. The best mapped points are (Tx,Rx) = (0.75,1.25) for KAN and MLP and (1.00,0.50) for linear control. Boundary minima are reported as best sampled allocations, not global optima.

**Stationary dwells separate convergence from deformable-filter advantage**

To determine whether the deformable-filter advantage could be explained by insufficient convergence time for the conventional adaptive equaliser, we presented five channel conditions in randomised order and held each stationary for 12, 24, or 48 consecutive HIL rounds (Fig. 5A and Supplementary Fig. S8). Receiver and FSE states persisted across the complete sequence, including dwell boundaries, and neither algorithm received condition identity.

The longer stationary intervals allowed the FSE to improve substantially within every condition (Fig. 5A). Comparing the first and final quarters of each dwell, FSE EVM decreased by 13.3% under combined bandwidth-nonlinear distortion at 0.55/0.769, by 8.3% under bandwidth limitation at 0.55, by 20.2% under combined distortion at 0.625/0.62, by 18.9% under bandwidth limitation at 0.70, and by 5.7% under the most severe combined condition at 0.475/0.90. The conventional equaliser was therefore not simply being denied sufficient adaptation time: under stationary conditions it continued to converge materially, in some cases by almost 20% over the course of a single dwell.

Despite this convergence, deformable filtering generally retained an advantage in the later part of each dwell (Fig. 5B). Using the final quarter of each stationary interval, the two-sided KAN system reduced EVM relative to the converged FSE by approximately 28.4%, 18.5%, 18.7%, 22.8%, and 13.6% across the five conditions, respectively. The corresponding MLP improvements were approximately 29.0%, 20.6%, 21.4%, 27.6%, and 11.7%. The linear controller followed the same trend in four conditions, with improvements of 38.4%, 39.7%, 21.5%, and 24.9%, but under the most severe combined condition its two-sided configuration finished approximately 2.6% worse than the FSE. This exception is important: deformable filtering did not produce a universal advantage independent of controller architecture, even when the underlying receiver structure was held fixed.

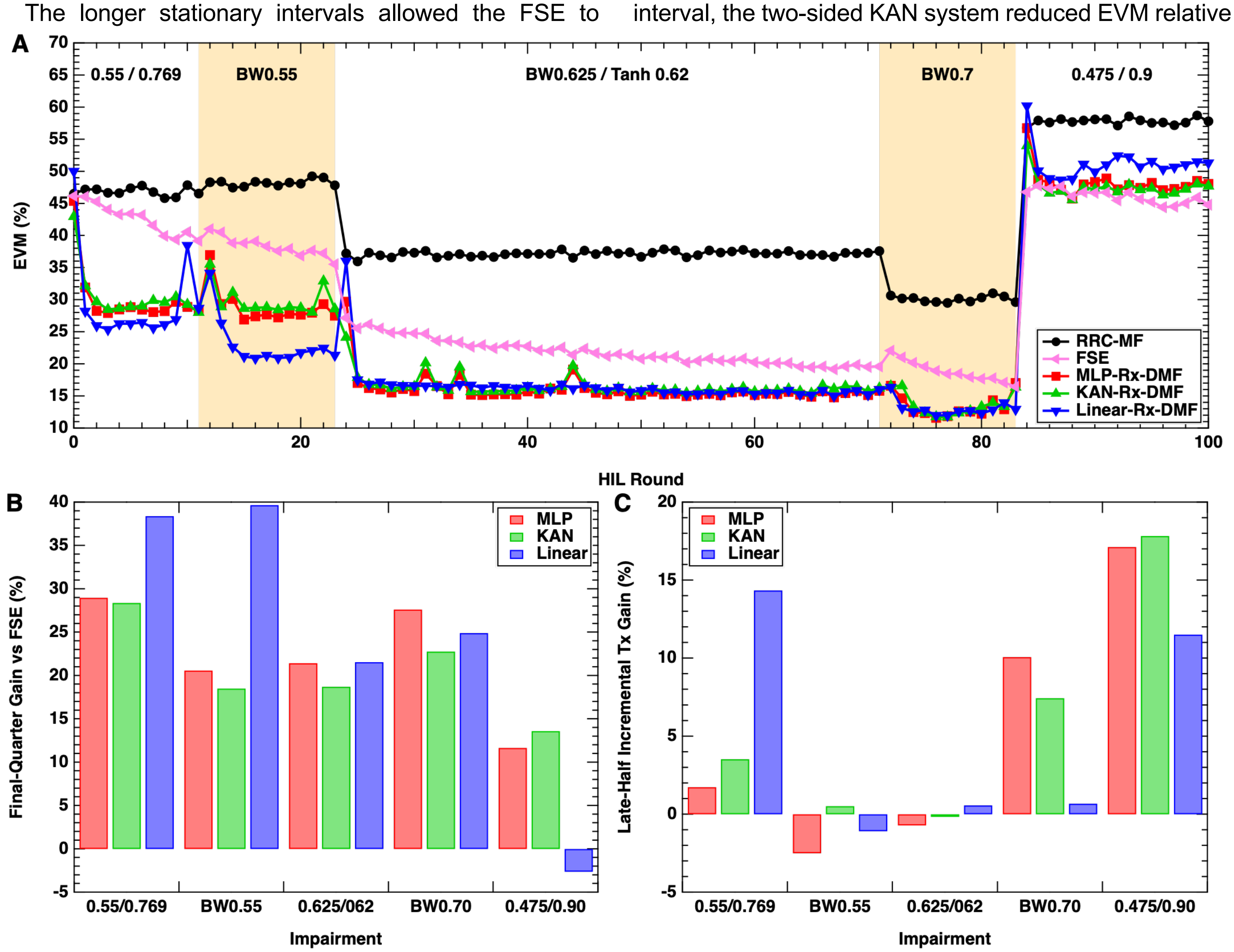

**Fig. 5. Stationary channel dwells separate conventional convergence from deformable-filter advantage.** (A) Receiver-only EVM trajectories during randomised stationary dwells of 12, 24, or 48 HIL rounds. Receiver and FSE states persist across all dwell boundaries, and condition identity is unavailable to adaptation. (B) Relative EVM reduction of each two-sided DMF system versus the FSE in the final quarter of each dwell. KAN and MLP remain ahead in all five conditions; linear control is ahead in four and slightly behind under the hardest combined distortion. (C) Incremental relative EVM reduction from adding the frozen transmitter deformation to the corresponding adaptive receiver, evaluated over the late half of each dwell. Bars are descriptive means over the prespecified windows. Combined-condition labels x/y denote bandwidth fraction x followed by nonlinear strength y.

The stationary dwells also clarified when transmitter deformation remained useful after the receiver had time to adapt (Fig. 5C). Under the most severe combined condition, adding the fixed transmitter deformation to the already adaptive receiver reduced late-half EVM by 17.8% for KAN, 17.1% for MLP, and 11.5% for the linear controller. By contrast, the incremental transmitter contribution was small or even slightly negative in several milder conditions. Thus, the complementarity observed in the earlier severity sweep persisted after extended receiver adaptation and became strongest when bandwidth restriction and nonlinear distortion were simultaneously severe.

These results resolve an important ambiguity raised by the adversarial rapid-switching stress test. The advantage of deformable filtering cannot be explained solely by faster response to rapidly changing conditions, because the conventional FSE was allowed to converge substantially under stationary operation and yet remained inferior to KAN- and MLP-controlled deformation in every tested dwell. At the same time, the linear controller's failure to remain ahead of the FSE in the hardest condition shows that controller structure still matters once the deformation problem becomes sufficiently demanding. The common benefit therefore lies in the deformable-filter formulation, while the complexity required of the controller depends on the channel regime.

**A full-day switching run tests endurance**

Finally, we tested the platform in a single uninterrupted 24.024-hour hardware session comprising 542 complete logical HIL rounds (Fig. 6A). Six channel epochs were applied for approximately four active hours each: bandwidth fractions 0.70 and 0.55; combined settings 0.625/0.62, 0.55/0.769, and 0.475/0.90; and a return to bandwidth fraction 0.70. Every round contained paired conventional and fixed Tx-DMF acquisitions, with their order alternated exactly (271 rounds in each order). Receiver and FSE states persisted independently for each transmitter arm. No condition identity, transition marker, payload reference, or scheduled reset entered the online algorithms, and no completed round or time interval was excluded.

The complete trajectories separated bounded adaptation from destructive state accumulation (Fig. 6A and Supplementary Figs. S10 and S11). During the hardest combined epoch, late-quarter EVM was 58.10% for the fixed matched-filter link and 74.84% for the persistent FSE, compared with 47.75% for KAN Rx-DMF, 47.94% for MLP Rx-DMF, and 48.42% for linear Rx-DMF. With the frozen transmitter deformation included, the corresponding KAN, MLP, and linear values were 42.12%, 42.35%, and 46.18%. The platform therefore remained bounded in the regime where accumulated FSE state became harmful, while controller-specific transients remained visible.

The return to bandwidth 0.70 provided a matched physical control for long-horizon recovery (Fig. 6B,C). Fixed-MF late-half EVM was essentially unchanged between the initial and returned visits (30.35% and 30.49%). Receiver-only KAN, MLP, and linear DMFs returned from 13.08%, 12.91%, and 11.42% to 11.90%, 12.06%, and 11.81%, respectively. The corresponding two-sided systems returned from 12.96%, 12.78%, and 11.79% to 12.29%, 12.33%, and 12.39%. The persistent FSE instead rose from 14.56% to 98.44% and continued to deteriorate after the physical condition had returned. Frequency responses of the receive DMFs show how all three controllers reshaped the analytical filter during the initial bandwidth-limited dwell and after the 20-hour return (Fig. 6D). The main passband form remains recognisable while controller-specific band-edge lift and attenuation notches evolve, making the learned physical compensation directly inspectable.

**Controller diversity is a platform property**

Across qualification, allocation, stationary-dwell, and endurance experiments, no controller dominated every signalling format or channel regime. Linear control was often strongest under bandwidth-dominated distortion; KAN performed particularly well under clean, nonlinear, and the hardest combined conditions; and MLP frequently tracked KAN while showing distinct condition-dependent advantages. Under the hardest stationary combined condition, linear control was the only two-sided DMF to finish behind the converged FSE.

All three controllers received the same blind descriptors and controlled the same bounded family of matched-filter deformations. Their differences therefore isolate the mapping from observed state to deformation rather than the waveform-processing structure. The shared gains support the deformable-filter platform; the changing rankings show that controller capacity should match the complexity of the operating regime.

This distinction is the central design rule. The learning problem is not to rediscover symbol recovery from sampled data, but to estimate how a known analytical solution should move as its assumptions fail. A linear map can be the most efficient controller when that relation is simple, whereas KAN or MLP control becomes valuable when interacting impairments require a nonlinear state-to-deformation map.

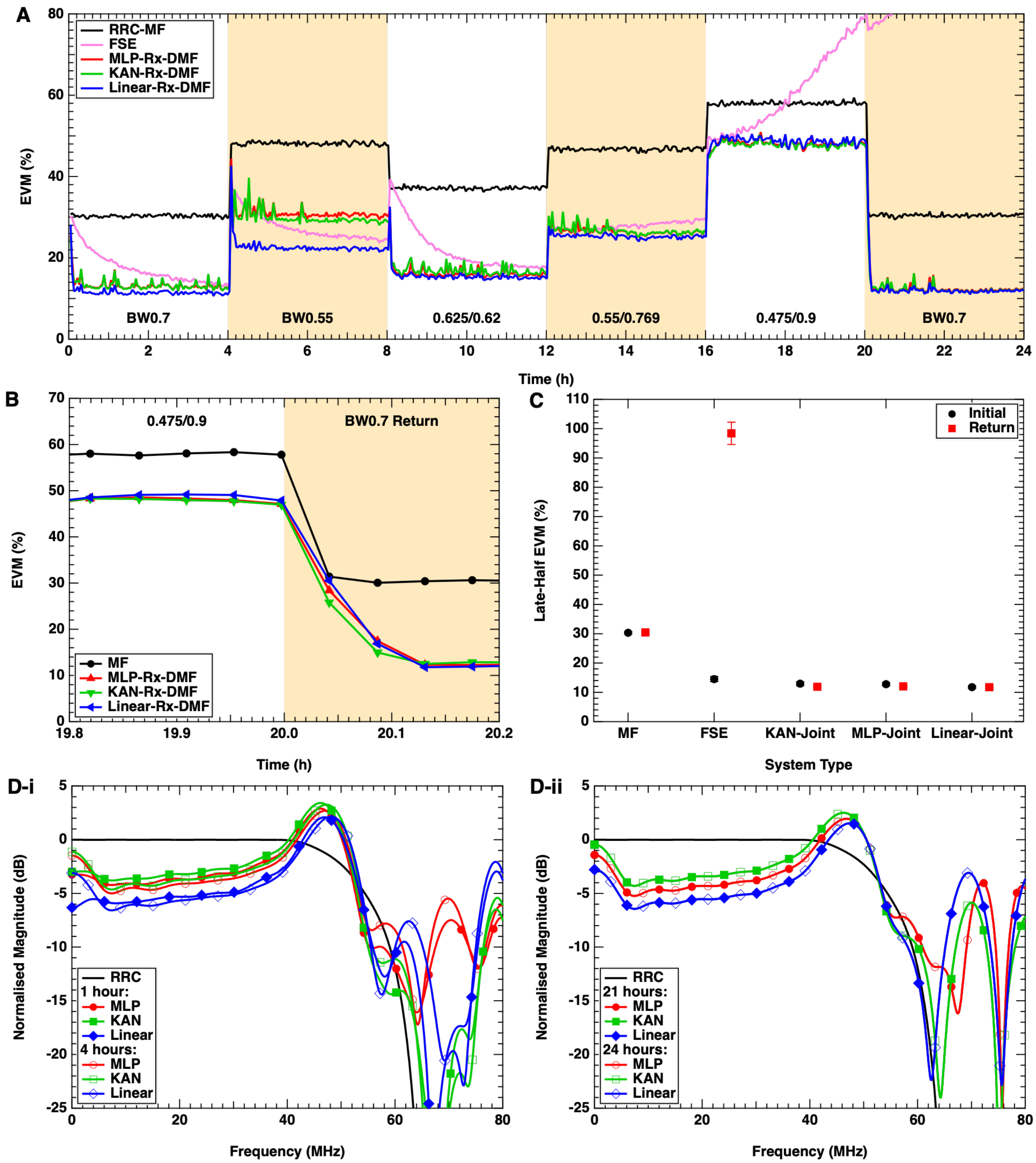

**Fig. 6. Bounded deformable receivers recover during uninterrupted 24-hour changing-condition operation.** (A) Complete receiver-only trajectories across six consecutive approximately four-hour physical channel epochs. The fixed matched filter, persistent span-matched FSE, and KAN-, MLP-, and linear-controlled Rx-DMFs use the conventional RRC transmitter. Controller traces are per-round means across four persistent seeds. (B) Transition from the hardest combined condition back to bandwidth fraction 0.70. The deformable receivers recover the original EVM regime, whereas the FSE continues to diverge. (C) Late-half EVM for the fixed MF, FSE, and two-sided DMF systems during the initial and returned visits to bandwidth fraction 0.70; error bars show descriptive standard deviations across 46 initial or 45 return HIL rounds. (D-i) Normalised magnitude responses of the receive DMFs after approximately 1 and 4 hours in the initial bandwidth-0.70 epoch. (D-ii) Responses after approximately 21 and 24 hours following return to the same physical condition. These are receiver filters from the RRC-transmitter arm, not transmitted pulses. The black curve is the canonical RRC matched-filter response. All 542 rounds were acquired in one 24.024-hour session without a scheduled receiver reset; condition labels and transmitted references were unavailable to the online controllers. Combined-condition labels x/y denote bandwidth fraction x followed by nonlinear strength y.

## DISCUSSION

These experiments establish deformable matched filtering as an adaptive platform rather than a monolithic learned receiver. The analytical filter remains the waveform-processing object, while learning is restricted to steering a bounded, inspectable deformation from one causal block to the next. The contribution is therefore not that KAN or another controller replaces a classical receiver; it is that a classical solution can learn continuously without surrendering its identity.

The transmitter and receiver experiments also exposed a useful asymmetry. Receiver deformation benefited strongly from continued online adaptation because the received waveform reflects the current channel, hardware state, and accumulated distortion. Transmitter deformation, by contrast, captured most of its benefit through a single low-dimensional pulse modification that generalised across held-out operating conditions, with only a small residual gain from continued transmitter adaptation. This suggests that transmitter and receiver deformation operate on different timescales: the transmitter can establish a broadly favourable waveform before propagation, whereas the receiver remains the natural location for rapid and persistent tracking of the realised link.

Two-sided operation did not produce a fixed additive gain. Under mild distortion, receiver adaptation alone captured most of the available improvement, and transmitter deformation contributed little once the receiver had adapted. As bandwidth limitation and nonlinearity increased, however, changing the waveform before propagation became progressively more useful. The joint-allocation experiment reinforced this point by showing that the preferred division of compensation depended on controller architecture and that maximising the available deformation at both ends was not universally beneficial. These results are consistent with a physical interpretation in which receiver-side processing can compensate distortions that remain recoverable after propagation, whereas transmitter shaping becomes more valuable when the channel itself suppresses or distorts information in a way that is difficult to undo retrospectively.

The comparison with the fractionally spaced equaliser is particularly important because it separates deformable matched filtering from a simple lack of conventional adaptation. The FSE was allowed to persist across rounds and, in the stationary-dwell experiment, improved materially within every condition. Deformable filtering nevertheless remained better in most late-dwell comparisons, including after the FSE had been given substantial time to converge. At the same time, the linear deformable receiver fell behind the FSE under the hardest combined condition. The experiments therefore do not support a claim that deformable matched filtering universally dominates conventional equalisation. Rather, they show that constraining adaptation around the matched-filter solution can remain advantageous even when a conventional adaptive filter is given the opportunity to converge, while also revealing regimes in which controller capacity becomes a limiting factor.

The controller comparison is a feature of the platform rather than a search for a universal winner. Because KAN, MLP, and linear control receive the same descriptors and act through the same filter family, changing rankings separate controller capacity from signal-processing structure. Linear control can be most effective when blind state maps simply onto the required deformation, whereas nonlinear models retain flexibility when impairments interact. The design rule is therefore to choose the explicit deformable object first and match controller capacity to the physical regime.

The architecture also changes the usual relationship between learning and the communication waveform. The controller never acts as a detector and never receives payload samples. Blind descriptors configure an explicit filter that processes only subsequent data. This removes pilots from ordinary adaptation, preserves a causal boundary between observation and action, and leaves the resulting physical filter directly inspectable in time or frequency.

The study has several limitations. The study establishes generalisation across held-out waveforms and impairment regimes within one optical wireless platform, not universal transfer across physical media. It uses a constrained family of bandwidth and memoryless nonlinear impairments. The adaptive interface itself is nevertheless not optical-specific: it requires a canonical reference filter, blind state descriptors derived from received samples, and a bounded deformation family, while no wavelength, optical-device variable, or channel label enters the controller. Radio-frequency, fibre, and acoustic implementations therefore provide direct tests of the same platform hypothesis rather than requiring a different algorithmic formulation. Demonstrating that cross-medium transfer, together with naturally drifting field channels, remains future work. The HIL loop establishes causal physical adaptation but is not yet a real-time embedded implementation; recent optical neural processors show that line-rate learned equalization is feasible in other architectures (17). The allocation grid is finite, and boundary minima do not identify global optima. The span-matched FSE is a strong same-stage benchmark but does not exhaust classical or learned equalizers. Finally, EVM is the continuous loss optimized in this study and resolves waveform-error changes that BER discards after hard decisions. Lower EVM does not by itself guarantee a fixed coded-link gain; achievable information rate and pre- and post-FEC performance remain necessary system endpoints.

Immediate extensions are independent hardware platforms, naturally drifting channels, higher-order and coded modulations, and FPGA implementations that close the loop at communication timescales. Regime-aware controller selection is also attractive, provided selection remains blind and preserves the common explicit filter structure. More broadly, the same principle can be applied to analytical signal-processing blocks whose optimality depends on assumptions that real systems only approximately satisfy.

The broader implication is that analytical and learned signal processing need not be treated as competing design philosophies. An analytical solution can instead define the structure, reference state, and physical constraints of an adaptive system, while learning determines how that solution should move as its assumptions become imperfect. In this formulation, learning does not replace communication theory; it operates around it.

## MATERIALS AND METHODS

### Study objectives and design

We designed the experiments to test whether an analytically defined communication filter could remain an explicit part of the signal-processing chain while learning continuously from blind observations of the received signal. The study was organised as a sequence of development, qualification, composition, allocation, and persistence experiments. Components were fixed before each subsequent qualification stage so that later results reflected adaptation within a prespecified architecture rather than repeated redesign of the receiver.

Throughout the study, the deformable matched filter (DMF) began from the conventional matched-filter solution and was constrained to remain within a bounded family of deformations around that reference. Learning did not operate directly on the payload waveform. Instead, each received block was summarised by blind receiver-state descriptors, which were supplied to a controller that determined the deformation parameters applied to subsequent blocks. No transmitted payload reference, pilot sequence, impairment label, or modulation-specific control signal was available to the controller during online adaptation. Reference symbols were retained only for offline performance scoring.

We enforced strict causality at the block level. A hardware-in-the-loop (HIL) round comprised one complete interaction with the experimental link: generation and transmission of a new waveform, physical acquisition of the received signal, receiver processing and offline performance scoring, followed by a causal update of the adaptive receiver state. Observations obtained during round could therefore influence the receiver only from round onward. Unless a receiver-declared safety mechanism triggered a reset or reheat, controller weights and deformable-filter state persisted across successive rounds.

The term *prespecified receiver* refers throughout to a receiver whose architecture, blind feature set, hyperparameters, deformation constraints, adaptation policy, and stability controls were fixed before evaluation. These design choices were not modified during the corresponding qualification experiment, but the online controller weights and filter state remained free to evolve causally during transmission. By contrast, a *fixed matched filter* denotes a receiver with no online adaptation, and a *fixed transmitter deformation* denotes a transmit pulse whose deformation coefficients remained constant during operation.

We evaluated three controller classes within the same deformable-receiver structure: a Kolmogorov-Arnold network (KAN), a multilayer perceptron (MLP), and a linear mapping. Each controller received the same blind state descriptors and controlled the same allowed filter family. This separation allowed the effect of the deformable-filter architecture to be distinguished from the expressive capacity of the controller itself.

As conventional references, we used the nominal matched-filter receiver and a span-matched fractionally spaced adaptive equaliser (FSE). The FSE operated on the oversampled waveform before symbol decisions and therefore provided a conventional adaptive-filter benchmark at the same stage of the receiver chain as the deformable filter. An in-line neural equaliser was not used as the principal benchmark because it would directly transform the payload waveform and would therefore change both the signal-processing architecture and the role assigned to learning.

The experimental sequence progressed from receiver-only development and held-out qualification to transmitter deformation, two-sided composition, deformation-budget allocation, a supplementary adversarial switching control, randomised stationary-dwell testing, and 24-hour endurance operation. Later experiments reused prespecified transmitter pulses and receiver designs without architecture-specific retuning unless the purpose was explicitly to vary the allowed deformation budget. This staged design separated generalisation, complementarity, allocation, convergence, and endurance rather than combining them into a single optimisation problem.

### Optical wireless hardware

We evaluated the framework using a hardware-in-the-loop optical wireless communication link. Digital waveform output and acquisition used a Xilinx Zynq UltraScale+ MPSoC ZCU102 evaluation board carrying an Analog Devices AD-FMCDAQ3-EBZ data-conversion module. The DAQ3 module contains an AD9152 16-bit digital-to-analogue converter and an AD9680 14-bit analogue-to-digital converter. The optical transmitter was an OPV300 series 850-nm vertical-cavity surface-emitting laser operated at a 6.5-mA drive current using Thorlabs laser-control electronics. The free-space distance from transmitter to receiver was 200 mm. Optical power was detected by a Newport photodetector 818-BB-21A and returned to the DAQ3 analogue-to-digital input. Adaptive processing ran between successive physical acquisitions; the FPGA and data-conversion hardware provided the repeatable waveform-generation and capture interface rather than a claim of real-time embedded controller inference. The same transmitter, optical path, receiver front end, and acquisition hardware were used throughout all comparative experiments.

The transmitted waveform was generated at 12 samples per symbol and passed through the physical transmitter and receiver chain before digital acquisition. Each acquisition contained multiple independently generated symbol blocks, allowing receiver-state descriptors, adaptation updates, and performance metrics to be evaluated repeatedly from fresh hardware measurements. Unless explicitly stated otherwise, the same acquisition was reused when comparing receiver structures that operated on an identical transmitted waveform. This paired design reduced the influence of acquisition-to-acquisition hardware variation when comparing the fixed matched filter, deformable receivers, and fractionally spaced equaliser.

The experiments were conducted as repeated physical acquisitions rather than as a software channel simulation. Bandwidth limitation and nonlinear distortion were imposed on the waveform before physical transmission, after which the resulting signal still propagated through the complete experimental hardware chain. Consequently, every reported HIL observation included the effects of the imposed test condition together with the residual bandwidth, noise, timing variation, and hardware imperfections of the optical link itself.

The acquisition procedure included frame synchronisation and quality checks before a capture was accepted for receiver processing. Captures that failed the acquisition-validity criteria were rejected rather than repaired using transmitted-symbol information. The adaptive receiver therefore operated only on physically acquired waveforms that satisfied the same acquisition contract, while payload references remained outside the online adaptation path and were used only for subsequent performance evaluation.

**Waveform generation**

We generated binary pulse-amplitude modulation (PAM-2) and partial-response PR4 waveforms to test whether the same adaptive receiver structure remained useful across different temporal signal relationships. PR4 signalling used the partial-response polynomial $1 - D^2$ corresponding to taps [1, 0, -1]. The adaptive controller was not informed which signalling format was present; signalling identity entered only the offline experimental organisation and performance analysis.

The nominal transmit waveform used root-raised-cosine (RRC) pulse shaping. Receiver qualification included previously unseen RRC roll-off factors of 0.15, 0.25, and 0.35, while the main persistent-link experiments used the common nominal roll-off of 0.25. Pulse shaping and partial-response generation were completed before physical transmission, and the resulting waveform was normalised according to the same transmit-chain convention before entering the hardware.

For paired system comparisons, payload generation was controlled so that the conventional and deformed transmitter arms used equivalent symbol realisations. This allowed changes produced by transmitter deformation to be separated from random differences in transmitted payload. The transmitted reference sequence was retained for offline EVM calculation but was never provided to the online receiver controller, blind feature extractor, FSE adaptation rule, or transmitter-feedback controller.

**Physical channel impairments**

We evaluated receiver and transmitter deformation under four broad channel regimes. Clean operation meant that no additional digital impairment was imposed. Bandwidth limitation used a finite-impulse-response low-pass filter whose bandwidth fraction specified the retained proportion of the nominal signal bandwidth. Nonlinear distortion used a memoryless hyperbolic-tangent transfer characteristic, parameterised by its strength. Combined conditions applied the low-pass filter and hyperbolic-tangent operation sequentially. The resulting waveform then passed through the complete optical hardware chain, so every measurement also contained the residual bandwidth, noise, timing variation, and hardware imperfections of the physical link. Throughout the text and figures, a compact combined-condition label x/y denotes bandwidth fraction x followed by nonlinear strength y; for example, 0.625/0.62 means bandwidth fraction 0.625 followed by hyperbolic-tangent strength 0.62.

Receiver qualification (Fig. 1) and held-out frozen-transmitter evaluation (Fig. 2) used clean operation, bandwidth fraction 0.625, nonlinear strength 0.62, and combined condition 0.625/0.62. Static transmitter-pulse development used a deterministic five-condition cycle comprising clean operation, bandwidth fractions 0.70 and 0.55, nonlinear strength 0.769, and combined condition 0.70/0.769. The later severity experiments extended the bandwidth series through 0.70, 0.625, 0.55, and 0.475 and the nonlinear series through strengths 0.50, 0.62, 0.769, and 0.90, with combined conditions 0.625/0.62, 0.55/0.769, and 0.475/0.90. Lower bandwidth fractions and larger nonlinear strengths denote progressively stronger restriction.

The adaptive algorithms were not provided with the impairment class or parameter values. Condition identity was used only to organise the experimental protocol and subsequent analysis. In particular, transitions between bandwidth-limited, nonlinear, and combined regimes did not trigger architecture-specific controller changes or condition-dependent adaptation rules.

The persistence experiments deliberately used abrupt changes between impairment regimes. These transitions were designed as stress tests of state carryover rather than as statistical models of a naturally drifting field channel. Separate stationary-dwell experiments were therefore used to distinguish transient adaptation effects from performance after longer convergence within a fixed physical condition.

**Nominal matched filtering**

The conventional matched filter provided both the receiver baseline and the analytical reference around which the deformable receiver operated. For the nominal RRC transmitter, the receiver used the corresponding canonical RRC matched-filter response. This filter remained fixed for the conventional matched-filter baseline and defined the zero-deformation state for the adaptive receiver.

The deformable receiver preserved this analytical response explicitly. At HIL round $k$, the receive filter was written as:

$$\mathbf{h}_{\mathrm{Rx}} = \mathbf{h}_{\mathbf{0}} + \Delta\mathbf{h}_{\mathrm{Rx},k}$$

where $\mathbf{h}_{\mathbf{0}}$ is the conventional matched filter and $\Delta\mathbf{h}_{\mathbf{Rx},k}$ is the learned deformation. The controller therefore did not synthesize an unconstrained receiver from scratch. It selected a bounded displacement around the conventional solution within a prescribed low-dimensional filter family.

The current received block was always processed using the filter state established before that block. Blind observations extracted from block $k$ could modify the deformation only for block $k + 1$ and later. This ordering prevented information derived from a block from being used to alter the filter that had already processed that same block.

**Receiver-state features**

The controller did not receive the sampled payload waveform directly. Instead, each acquisition was compressed into a fixed set of blind receiver-state descriptors designed to summarise properties of the currently observed signal without requiring transmitted symbols or channel labels.

The final receiver used 26 descriptors: six scalar signal-state measures, two moment descriptors, six short-lag autocorrelation terms, and twelve polyphase-energy descriptors (Supplementary Fig. S2). Together they summarise received-signal shape, temporal memory, and sampling-phase structure without signalling labels or explicit impairment identity.

The same 26-dimensional feature vector was supplied to the KAN, MLP, and linear controllers. Feature definitions were fixed before the held-out receiver qualification experiments and were not changed for different pulse shapes, signalling families, bandwidth restrictions, nonlinearities, or combined impairments. This ensured that subsequent performance differences reflected the controller mapping and adaptive state rather than condition-specific feature engineering.

Because the feature extractor operated only on blind observations of the received signal, the controller's task was fundamentally different from direct learned equalisation. It estimated how the explicit matched-filter response should change; it did not infer transmitted symbols from the waveform itself.

**Receiver deformation model**

KAN, MLP, and linear controllers mapped the same receiver-state vector onto parameters defining a deformation of the nominal matched filter. All three architectures therefore shared the same signal-processing path and differed only in the function used to relate blind receiver state to allowed filter displacement.

The 301-tap receive filter operated at 12 samples per symbol and was deformed using 64 smooth tapered discrete-cosine modes (Supplementary Fig. S1). The deformation magnitude was bounded to prevent arbitrary departure from the analytical reference. The same deformation family and stability budget were used for KAN, MLP, and linear control so that controller comparisons did not change the physical filter class.

The KAN and MLP controllers provided nonlinear mappings from receiver state to deformation parameters, whereas the linear controller supplied an affine mapping within the same deformation space. This common output representation was essential to the experimental design: the controller architecture could change the way receiver state was interpreted, but not the physical form of the filter that ultimately processed the waveform.

**Causal online adaptation**

Online receiver adaptation proceeded block by block and persisted across successive HIL rounds. At the start of a logical link, the receiver began from the conventional matched-filter state. Blind features extracted from each processed block were supplied to the controller, and the resulting state update affected only later blocks. Controller weights and deformable-filter state were preserved between HIL rounds unless the receiver's blind safety logic declared that a reset or reheat was required.

No pilot sequence, known training block, transmitted payload reference, modulation label, or impairment label entered this adaptation loop. Reference symbols were retained only outside the online receiver for performance scoring after processing. The receiver therefore had to infer the direction of useful deformation solely from its own blind state measurements.

This persistent operating model differs from evaluating an independently optimised filter at each channel condition. When the physical condition changed, the receiver entered the new regime with the state accumulated under the preceding conditions. The changing-condition and stationary-dwell experiments deliberately retained this history so that adaptation stability, convergence, and path dependence could be observed directly.

**Receiver stability controls**

Persistent online adaptation creates the possibility of gradual drift away from the useful matched-filter neighbourhood. We therefore incorporated blind stability mechanisms into the receiver before the final qualification experiments and retained them unchanged thereafter.

Following link initialisation or a receiver-declared matched-filter reset, the controller was allowed 24 adaptation blocks of unconstrained acquisition. The subsequent trust radius was calibrated independently for each controller and seed as 12 times the median root-mean-square magnitude of its first eight whole-block parameter updates. If accumulated displacement exceeded this radius, the parameter state was projected onto the allowed boundary. A moderate reheat reopened a 24-block grace interval while preserving the existing anchor and calibration; a matched-filter-class reset established a new anchor and calibration.

A separate blind health path monitored whether the current matched-filter-centred state had become unreliable. This mechanism remained active even after the receiver had adapted away from the nominal filter and could trigger a reheat or return toward the matched-filter reference without access to transmitted symbols. Scheduled condition changes did not themselves trigger these actions; they arose only from the receiver's blind internal criteria.

These controls were fixed before the held-out and persistent receiver experiments. Thus, later changes in performance reflected the evolution of the online state within a prespecified stability policy rather than manual intervention or condition-specific retuning.

**Static transmitter deformation**

We constrained transmitter adaptation to a low-dimensional deformation of the nominal RRC pulse rather than allowing arbitrary waveform synthesis. The transmit filter was written as:

$$\mathbf{h}_{\mathrm{Tx}} = \mathbf{h}_{\mathrm{RRC}} + \mathbf{B}_{\mathrm{Tx}}\mathbf{c}$$

where $\mathbf{B}_{\mathrm{Tx}}$ contained eight smooth tapered discrete-cosine basis functions and **c** contained their coefficients. The transmit filter comprised 193 taps, and the deformation norm was bounded so that the learned pulse remained a controlled perturbation of the conventional RRC response.

We first optimised a single static coefficient vector over a development set spanning PAM-2 and PR4 signalling and clean, bandwidth-limited, nonlinear, and combined impairment regimes. The resulting coefficient vector was then fixed for all subsequent transmitter qualification and two-sided experiments. Thus, *static transmitter deformation* means that the transmit pulse itself no longer changed online; it does not imply that the receiver was fixed when this pulse was later combined with adaptive reception.

The frozen coefficient vector was:

$$[-0.06756, -0.08992, 0.06913, 0.02501, \ldots -0.03848, -0.00869, 0.02937, -0.02219]$$

The resulting pulse was normalised using the same transmit convention as the nominal RRC pulse and was required to satisfy the same physical-validity constraints used during development, including bounded deformation magnitude and restrictions on residual tail energy, out-of-band energy, and peak amplitude.

Held-out transmitter qualification was performed at RRC roll-offs of 0.15, 0.25, and 0.35 without modifying the coefficients. These tests therefore measured transfer of the same physical pulse deformation rather than re-optimisation for each waveform.

**Delayed transmitter adaptation**

We separately tested whether continued transmitter adaptation could add useful residual improvement beyond the static pulse. In this experiment, the static transmit deformation served as the operating point and a small additional deformation was controlled from delayed blind feedback generated at the receiver.

The feedback vector contained six matched-filter memory descriptors:

$$[\rho_{MF}(1), \ldots, \rho_{MF}(6)]$$

selected during the preceding transmitter-feature study. These descriptors were computed from the received signal without transmitted-symbol references or channel labels. Feedback obtained after HIL round was available only for the transmitter configuration used from round onward.

KAN, MLP, and linear controllers were evaluated within the same residual transmitter-deformation space. Their model parameters had been determined before the online experiment; during operation, the controllers mapped the delayed six-dimensional receiver state to a causal residual adjustment of the transmitted pulse. This design prevented same-round information from affecting the waveform that generated that information.

Because the adaptive residual produced only a small additional gain beyond the static deformation, subsequent two-sided experiments used the static transmitter pulse. This kept the transmitter contribution fixed while allowing receiver adaptation and transmitter-receiver complementarity to be studied without introducing a second rapidly varying adaptive loop.

**Deformation allocation**

To study how compensation should be distributed between transmitter and receiver, we varied only the permitted deformation budgets while leaving the underlying transmitter direction and receiver adaptation machinery unchanged.

The transmitter deformation was scaled by:

$$\lambda_{\mathrm{Tx}} \in \{0, 0.25, 0.5, 0.75, 1, 1.25\}$$

where $\lambda_{\mathrm{Tx}} = 0$ restored the conventional RRC pulse and $\lambda_{\mathrm{Tx}} = 1$ reproduced the independently determined static transmitter deformation.

The receiver budget was scaled using:

$$\lambda_{\mathrm{Rx}} \in \{0.5, 0.75, 1, 1.25\}$$

This factor scaled the maximum allowed receive-filter deformation and the late-trust displacement radius. All other receiver components, including blind features, controller architecture, learning rules, safety logic, and causal update structure, were left unchanged. Receiver weights and filter state continued to adapt online within the selected budget.

The allocation experiment used three combined bandwidth-nonlinear conditions: 0.625/0.62, 0.55/0.769, and 0.475/0.90. For each physical transmitter setting, the same acquired waveform was reused across the four receiver budgets. This produced a paired two-dimensional response surface while avoiding unnecessary differences in physical channel realisation along the receiver-budget axis.

The point $(\lambda_{\mathrm{Tx}}, \lambda_{\mathrm{Rx}}) = (1, 1)$ served as the common reference used in the surrounding experiments. We report the lowest sampled points as *best mapped* allocations rather than global optima because the tested grid was finite and, for some controllers, the lowest EVM occurred at a boundary of the sampled receiver range.

**Fractionally spaced equaliser**

We used a decision-directed normalised least-mean-squares fractionally spaced equaliser as the principal conventional adaptive-filter benchmark. The FSE operated at two samples per symbol with 51 taps, corresponding to a temporal span of 25 symbol intervals and therefore approximately matching the temporal reach of the 301-tap receive-filter path at 12 samples per symbol (6, 9).

The equaliser operated after the same canonical matched-filter front end used in the comparison receiver path and before hard symbol decisions. It was initialised from the nominal identity-centred state and adapted using blind decisions rather than transmitted-symbol references, following the conventional adaptive-equalisation framework (19). The update step size was 0.01 with zero leakage.

The FSE remained continuously adaptive across HIL rounds. After processing round *k*, its updated coefficient vector was retained and used to initialise round *k* + 1, where further decision-directed NLMS updates were performed. Thus, the equaliser did not restart from its nominal state at each acquisition; it accumulated adaptation history over the duration of a logical link. As with the deformable receiver, observations from the current block could influence only subsequent adaptive updates, and transmitted payload references were used only for offline EVM calculation.

This comparator was chosen because it represents a conventional adaptive filtering solution operating at the same receiver stage as the DMF. A direct neural waveform equaliser would instead replace the explicit filtering path and would therefore test a different architectural hypothesis.

**Persistent changing-condition experiment**

We evaluated long-term state persistence using a deliberately non-stationary HIL stress test. The experiment comprised 180 logical rounds in which channel conditions changed repeatedly while adaptive receiver and FSE states were carried forward continuously.

Five complete system configurations were evaluated: conventional RRC transmission with fixed matched filtering, static transmitter deformation with fixed matched filtering, conventional transmission with adaptive receiver deformation, combined static transmitter and adaptive receiver deformation, and conventional transmission with the span-matched FSE.

Each logical round used two physical acquisitions, one generated with the conventional transmitter and one with the static deformed transmitter. The conventional-transmitter acquisition was reused for matched-filter, deformable-receiver, and FSE processing. The deformed-transmitter acquisition was reused for transmitter-only and two-sided evaluation. This pairing enabled receiver structures to be compared using the same physical waveform realisation.

The receiver was not reset at condition transitions and received no indication that the impairment had changed. The protocol was intentionally more abrupt than a typical field channel and was designed as a stress test of persistent adaptation rather than as a statistical model of ordinary communications traffic.

**Randomised stationary-dwell experiment**

We performed a second persistence experiment to determine whether the changing-condition results could be explained by insufficient convergence time for the conventional adaptive equaliser. Five channel conditions were therefore presented in randomised order and held stationary for dwell lengths of 12, 24, or 48 consecutive HIL rounds.

The realised schedule was: combined 0.55/0.769 for 12 rounds, bandwidth 0.55 for 12 rounds, combined 0.625/0.62 for 48 rounds, bandwidth 0.70 for 12 rounds, and combined 0.475/0.90 for 24 rounds, giving 108 logical rounds in total.

The adaptive receiver and FSE state persisted throughout the complete 108-round sequence. Neither algorithm was reset on entry to a new dwell, and condition identity was not supplied to the adaptation mechanism. The same five link configurations and paired physical-acquisition protocol used in the changing-condition experiment were retained.

Within each dwell, we compared early and late portions of the trajectory to distinguish transient response from longer-term convergence. In particular, first-quarter and final-quarter EVM were used to quantify FSE convergence, while late-quarter and late-half statistics were used for comparisons between the FSE, receiver-only deformation, and two-sided deformation.

**Twenty-four-hour switching experiment**

We conducted a single-session 24-hour persistence experiment from 19 August 2026 06:32:53 UTC to 20 August 2026 06:34:19 UTC. The run accumulated 24.024 hours of active operation and contained 542 complete logical rounds with a mean duration of 159.57 s. Six conditions were scheduled by completed active time in nominal four-hour epochs: bandwidth 0.70, bandwidth 0.55, combined 0.625/0.62, combined 0.55/0.769, combined 0.475/0.90, and bandwidth 0.70 again. A condition was selected at the start of each logical round, so a single round could overrun a nominal boundary. Every round used paired physical acquisitions with the conventional RRC and frozen Tx-DMF pulses; acquisition order alternated between rounds to balance temporal ordering.

Adaptive receiver and FSE states persisted separately for the conventional and Tx-DMF transmitter arms across all 542 rounds and physical condition boundaries. No scheduled reset, pilot, training sequence, condition identity, or transition timing was supplied to adaptation; the predeclared blind change detector and safety interlocks remained active. Round indices 0 to 541 were contiguous, and no completed round was excluded. Controller trajectories are means across four persistent seeds. For the frequency-response snapshots in Fig. 6D, receive filters were reconstructed for the conventional-transmitter arm at the selected physical rounds nearest 1, 4, 21, and 24 hours. KAN, MLP, and linear responses were calculated from their persistent controller states and the measured first-block blind descriptors for the corresponding capture. Magnitudes were normalised to the canonical matched-filter reference and plotted against absolute digital frequency using the experimental sampling rate. No transmitter-filter response is included in this panel.

**Error-vector magnitude**

We used error-vector magnitude (EVM) as the primary performance metric because it is the continuous waveform-error objective against which the transmitter and receiver designs were optimised and compared. Unlike bit-error rate, which thresholds recovered symbols into correct or incorrect decisions and can therefore remain unchanged across materially different residual distortions, EVM retains the magnitude of the symbol error and resolves improvement throughout the operating range. EVM was calculated after receiver processing using known transmitted symbols retained outside the online adaptation loop. These references were never available to the deformable receiver, FSE, or transmitter-feedback controller during operation; they were used for offline optimisation, evaluation, and comparison only (18).

For each evaluated capture, the recovered symbol sequence was aligned with the corresponding transmitted reference, and EVM was computed from the residual complex or real-valued symbol error after the same offline scoring procedure had been applied to all receiver structures. EVM therefore measured distance from the ideal constellation before hard-decision thresholding. It should be interpreted as the study's system-level optimisation and evaluation objective, not as a reference signal exposed to the causal online controller.

For experiments containing both PAM-2 and PR4 signalling, we retained signalling-family-specific EVM values and, where a single summary value was required, averaged the PAM-2 and PR4 contributions with equal weight. We avoided using pooled means across heterogeneous channel conditions as the principal basis for interpretation; condition-resolved trajectories, impairment-specific comparisons, and within-dwell statistics were used wherever the channel state itself was central to the experimental question.

**Relative performance metrics**

For comparisons against a reference receiver, we expressed improvement as the fractional reduction in EVM:

$$G = \frac{\mathrm{EVM}_{\mathrm{ref}} - \mathrm{EVM}_{\mathrm{test}}}{\mathrm{EVM}_{\mathrm{ref}}}$$

with positive values indicating lower EVM for the tested system. Percentage improvements reported in the text and figures are $100G$.

To quantify the incremental value of transmitter deformation when the receiver was already adaptive, we used:

$$G_{\mathrm{Tx|Rx}} = \frac{\mathrm{EVM}_{\mathrm{Rx\text{-}only}} - \mathrm{EVM}_{\mathrm{Tx+Rx}}}{\mathrm{EVM}_{\mathrm{Rx\text{-}only}}}$$

This quantity isolates the additional contribution of changing the transmitted pulse from the larger receiver-adaptation benefit that may already be present.

For the stationary-dwell experiment, convergence was evaluated by comparing EVM over the first and final quarters of each dwell. Comparisons intended to represent settled behaviour used the final quarter, while transmitter–receiver complementarity was evaluated over the late half of the dwell to reduce sensitivity to the immediate transition transient.

**Statistical analysis**

Statistical summaries were matched to the experimental hierarchy and are primarily descriptive because controller seeds, waveform blocks, and successive persistent HIL rounds are repeated observations rather than independent hardware replicates. Receiver comparisons were paired whenever competing receivers processed the same physical acquisition.

For held-out transmitter qualification, each roll-off contained 48 matched RRC/Tx-DMF evaluations. EVM error bars show standard errors across these evaluations, and relative-gain intervals are two-sided 95% t-confidence intervals on the paired gain values. For controller comparisons, seed standard deviations describe variability among the four persistent controller realisations and do not increase the physical replicate count.

For the experiment repeats, seeds and captures were averaged within each run, roll-off, condition, HIL index, and receiver cluster. Error magnitudes were then calculated across the three roll-off clusters at each HIL index. The repeat correlations and adaptation-event plots are mechanistic diagnostics, not independent-sample hypothesis tests.

For stationary dwells, the first and final quarters quantify within-dwell FSE convergence. Final-quarter comparisons against the FSE and late-half transmitter increments use prespecified windows and are reported as effect sizes. The shortest 12-round dwells contribute only three rounds to a final-quarter estimate and are therefore interpreted through effect direction and consistency rather than narrow inferential intervals.

For the 24-hour experiment, all 542 completed rounds are shown. Late-quarter and late-half values are arithmetic means over prespecified contiguous windows. Error bars in the repeated-condition comparison are descriptive standard deviations across physical HIL rounds. Temporal dependence is preserved in the displayed trajectories; recorded round counts are not presented as independent sample sizes.

**Reproducibility and experimental control**

All controller architectures, feature definitions, deformation spaces, adaptation rules, safety policies, transmitter coefficients, and experimental schedules used for qualification were recorded in version-controlled configuration files before the corresponding evaluation stage. Later experiments reused these prespecified components unless the purpose of the experiment was explicitly to vary one of them, such as the transmitter or receiver deformation budget.

The receiver never received modulation identity, channel-condition identity, impairment severity, transmitted payload symbols, or pilot information as an online control input. Experiment metadata containing these quantities was retained only for hardware configuration, bookkeeping, and offline analysis. This separation was maintained throughout the HIL pipeline so that condition information used to construct an experiment could not become an implicit feature available to the adaptive controller.

For persistent experiments, adaptive state was checkpointed between successive HIL rounds so that the same logical link could continue from its previous state. Hardware or acquisition failures were treated as experimental interruptions rather than as opportunities to modify receiver behaviour. Where a run was resumed after an interruption, the resulting continuity and active operating time were recorded explicitly.

**Use of generative artificial intelligence**

OpenAI ChatGPT (GPT-5.6 Sol) was used during the research workflow to assist with code development and with computational inspection of experimental outputs. AI-assisted code and analyses were reviewed by the author, and all numerical results reported in the manuscript were verified against the underlying experimental data and analysis outputs manually in the first and last cases. The AI system did not generate experimental measurements, control the experimental hardware autonomously, determine which observations were included or excluded, or replace the prespecified adaptive algorithms evaluated in the study.

**Adversarial rapid-switching stress test (Supplementary)**

As a supplementary robustness control, we evaluated state persistence using an intentionally adversarial 180-round HIL sequence in which channel conditions changed repeatedly, often before any condition-specific steady state could be established. Adaptive receiver and FSE states were carried forward continuously. This experiment was designed to test bounded operation under hostile state transitions, not convergence, realistic channel dynamics, or condition-resolved performance.

Five complete system configurations were evaluated: conventional RRC transmission with fixed matched filtering, static transmitter deformation with fixed matched filtering, conventional transmission with adaptive receiver deformation, combined static transmitter and adaptive receiver deformation, and conventional transmission with the span-matched FSE.

Each logical round used two physical acquisitions, one generated with the conventional transmitter and one with the static deformed transmitter. The conventional-transmitter acquisition was reused for matched-filter, deformable-receiver, and FSE processing. The deformed-transmitter acquisition was reused for transmitter-only and two-sided evaluation. This pairing enabled receiver structures to be compared using the same physical waveform realisation.

**Funding**
This work did not receive financial support.

**Author contributions**
The author conceived the study, developed the methodology and software, performed the experiments and analysis, curated the data, prepared the visualisations, and wrote and revised the manuscript.

**Competing interests**
The authors declare that they have no competing interests.

**Data, code and materials availability**
All data required to evaluate the conclusions of the paper are available in the paper and/or the Supplementary Materials. The underlying hardware-in-the-loop measurements, processed result tables, configuration files, and analysis outputs supporting the figures will be deposited in https://github.com/qmul-optocomms/dmf-design-rules-public with a persistent identifier before publication. Source code required to reproduce the signal generation, deformable-filter processing, conventional receiver comparisons, statistical analysis, and figure generation will be made available in the same repository.

**Supplementary Materials for**
**Learning to deform the matched filter**

Paul Anthony Haigh
*Corresponding author. Email: p.a.haigh@qmul.ac.uk

**This PDF file includes:**

Supplementary Text
Figs. S1 to S12

## Supplementary Text

### Deformation geometry and blind state

The deformable receiver retains the canonical 301-tap matched filter at 12 samples per symbol as its exact zero state. Controllers act only through a frozen family of 64 smooth tapered modes (Fig. S1). Every controller receives the same 26 blind descriptors, divided into scalar, moment, autocorrelation, and polyphase groups (Fig. S2). This common input and output contract separates controller capacity from waveform-processing structure.

### Independent repeat and adaptation-state diagnostics

The partial experiment repeat used newly acquired physical waveforms while retaining the prespecified receiver architecture and schedule. Seeds and waveform blocks are repeated algorithmic observations, not independent hardware replicates. Clean-condition trajectories reproduce across the two acquisitions (Fig. S3), and the condition-resolved summary demonstrates changing controller advantages (Fig. S4). Small excursions coincide with the first activation of the late-trust constraint after its 24-block grace period (Fig. S9). This timing is consistent with deterministic projection and subsequent readaptation; no matched-capture no-update counterfactual is claimed.

### Transmitter controls and factorial boundary condition

The same frozen transmitter deformation is tested at all three held-out roll-offs and all 144 paired physical evaluations (Fig. S5). Its large transferable benefit contrasts with the much smaller residual gain from online transmitter adjustment. The factorial experiment (Fig. S6) establishes that this pulse is largely redundant after receiver adaptation under moderate distortion, motivating the severity sweep in main Fig. 3.

### Persistence controls

The rapid-switching sequence intentionally changes impairment nearly every HIL round and is used only to test bounded state carry-over (Fig. S7). The randomised stationary-dwell experiment provides the convergence control used for the main performance interpretation (Fig. S8). Full receiver-only, two-sided, and baseline trajectories from the uninterrupted 24-hour run, condition-resolved late-window performance, and a representative internal state trace are provided in Figs. S10 to S12.

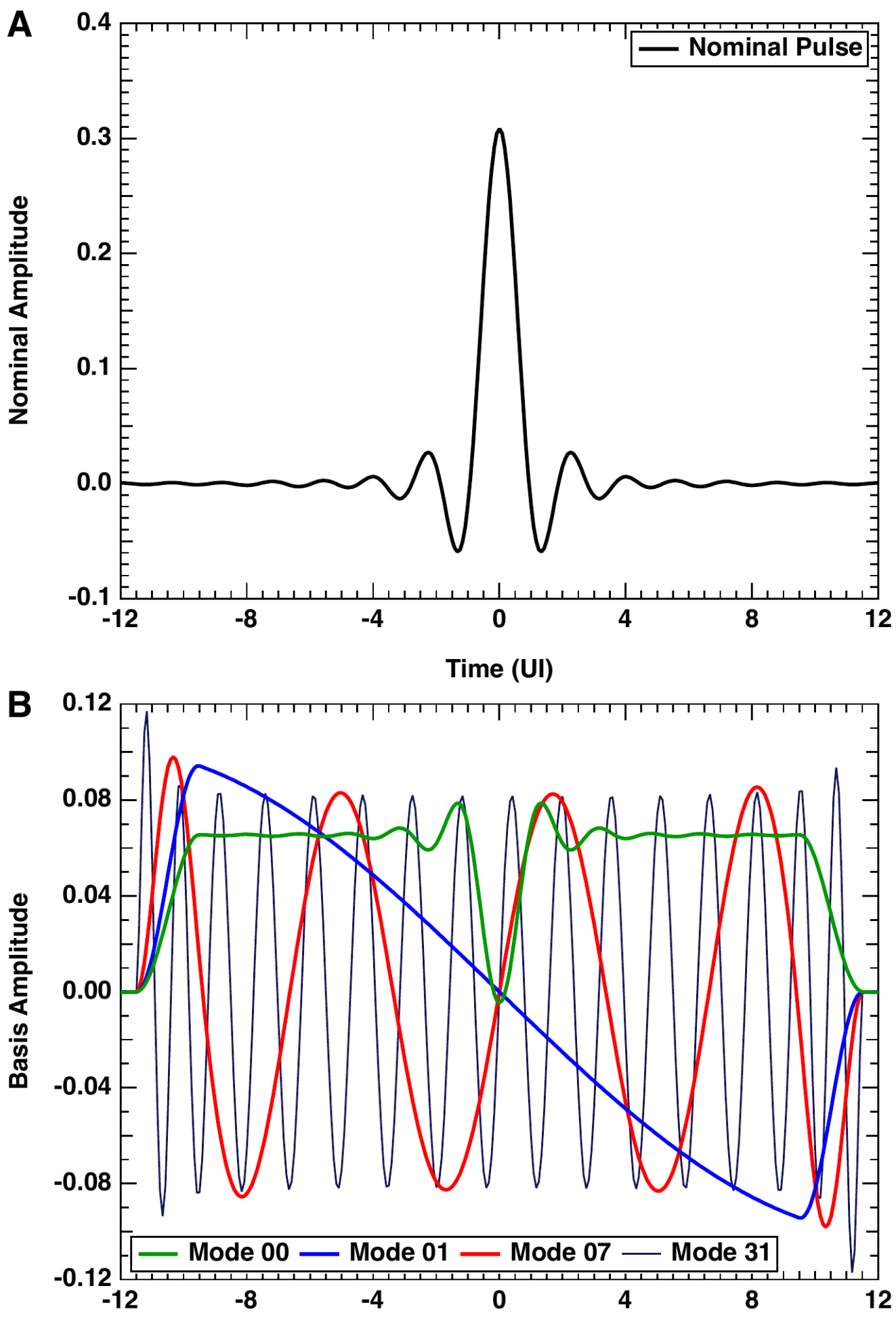

**Fig. S1. Matched-filter deformation geometry.** (A) Canonical 301-tap RRC matched-filter response at 12 samples per symbol, shown against time in unit intervals. (B) Four representative modes from the frozen 64-mode smooth tapered discrete-cosine deformation family. The zero-deformation state is exactly the response in (A); controllers select bounded mode amplitudes rather than synthesizing an unconstrained waveform filter

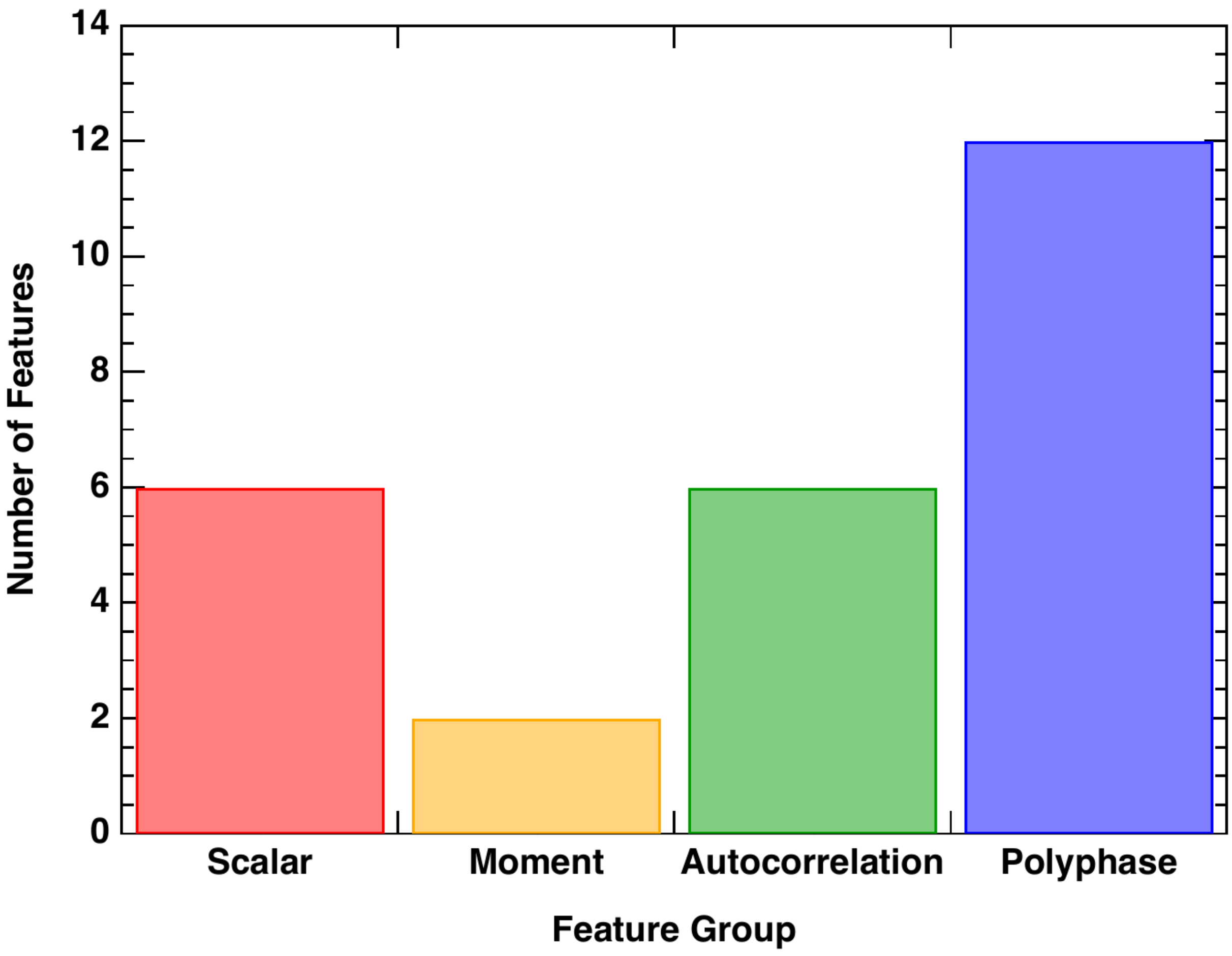

**Fig. S2. Frozen blind receiver-state vector.** Composition of the 26-element state supplied identically to KAN, MLP, and linear controllers: six scalar descriptors, two moment descriptors, six short-lag autocorrelation descriptors, and twelve polyphase-energy descriptors. No payload samples, transmitted symbols, modulation labels, or impairment labels are included.

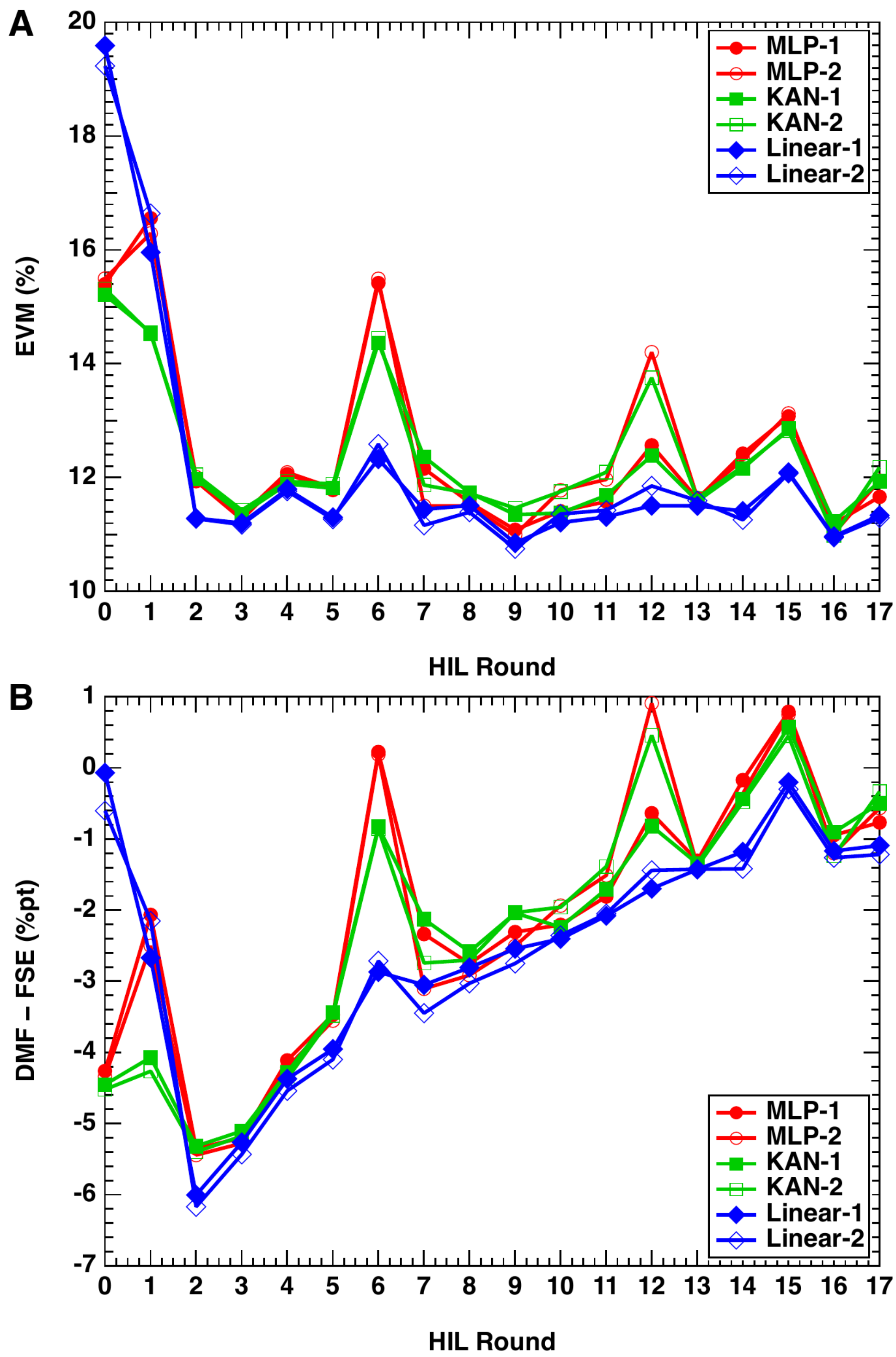

**Fig. S3. Independent physical repeat of clean-condition receiver qualification.** (A) Mean EVM trajectories for KAN, MLP, and linear Rx-DMFs in the first experiment and the independently acquired experiment repeat. (B) Corresponding EVM difference from the matched persistent FSE; negative values indicate lower EVM for the DMF. At each HIL index, controller seeds and captures are first averaged within each run and roll-off. Error bars are 95% t-confidence half-widths across the three roll-off clusters. Seeds and waveform blocks are repeated measurements, not independent hardware replicates.

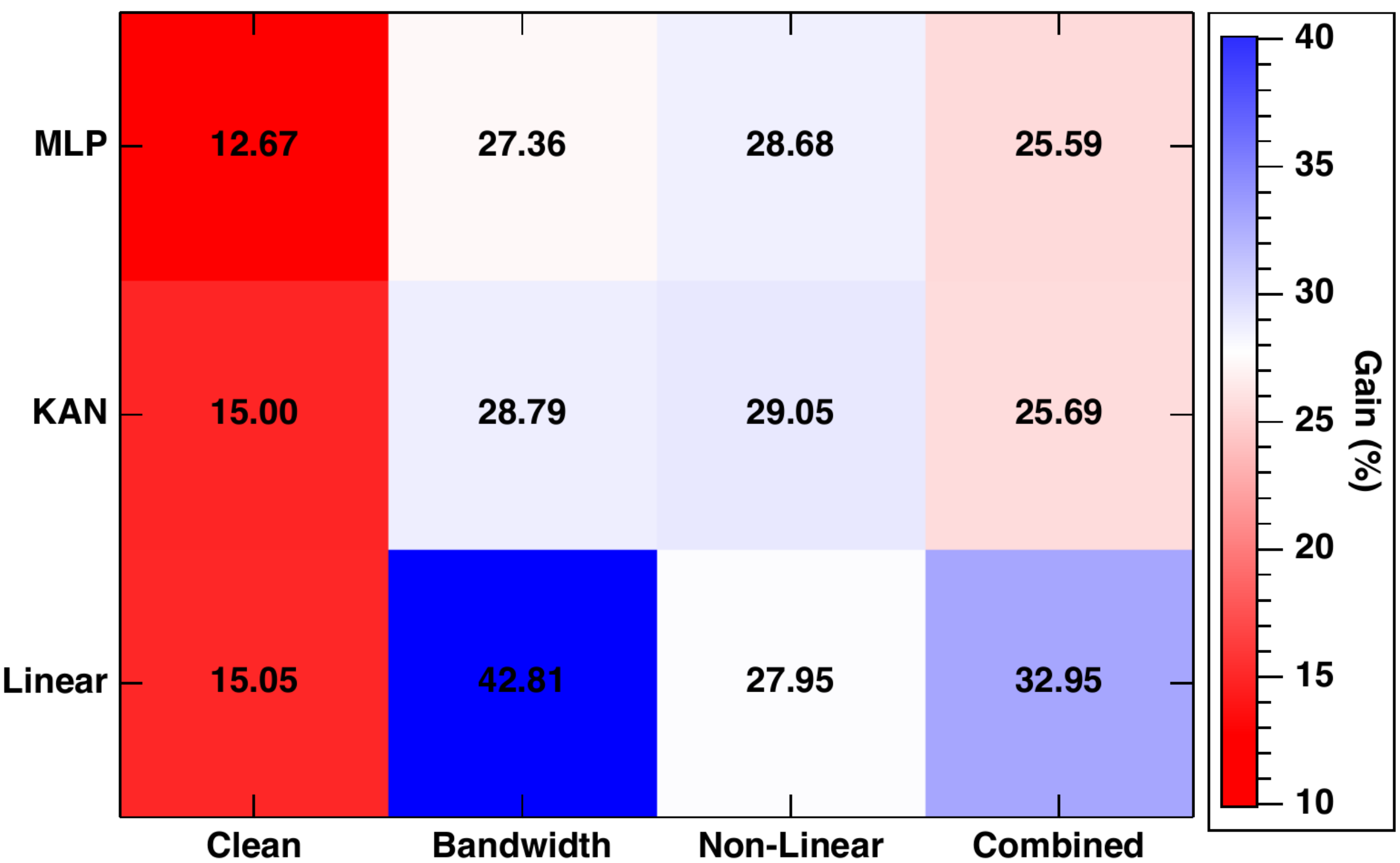

**Fig. S4. Controller-dependent operating advantages within one receiver structure.** Condition-resolved relative EVM reduction versus the persistent FSE during C12 qualification. Values are descriptive means across the stored matched capture, controller-seed, and block observations for each impairment class. The changing pattern across KAN, MLP, and linear control is the relevant result; the underlying repeated observations are not treated as independent physical replicates.

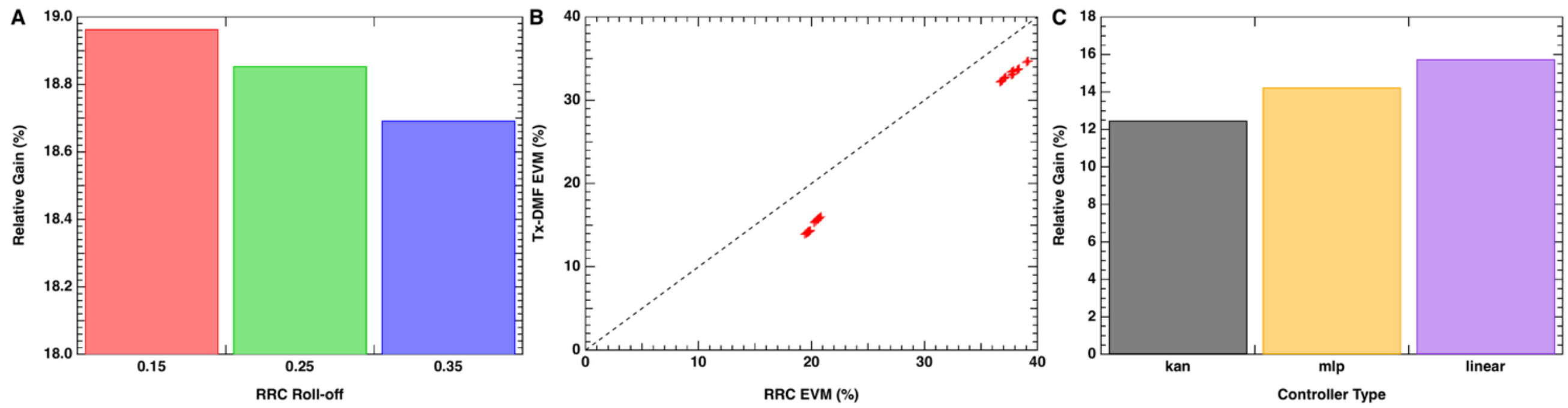

**Fig. S5. Held-out transmitter qualification and residual online adaptation.** (A) Mean relative EVM reduction of the frozen Tx-DMF pulse at held-out roll-offs of 0.15, 0.25, and 0.35. (B) Paired Tx-DMF and conventional RRC EVM for all 144 held-out evaluations; every point lies below the line of equality. (C) Incremental relative EVM reduction produced by continued delayed transmitter adaptation beyond the frozen pulse. Bars show means across four controller seeds and error bars show seed standard deviations.

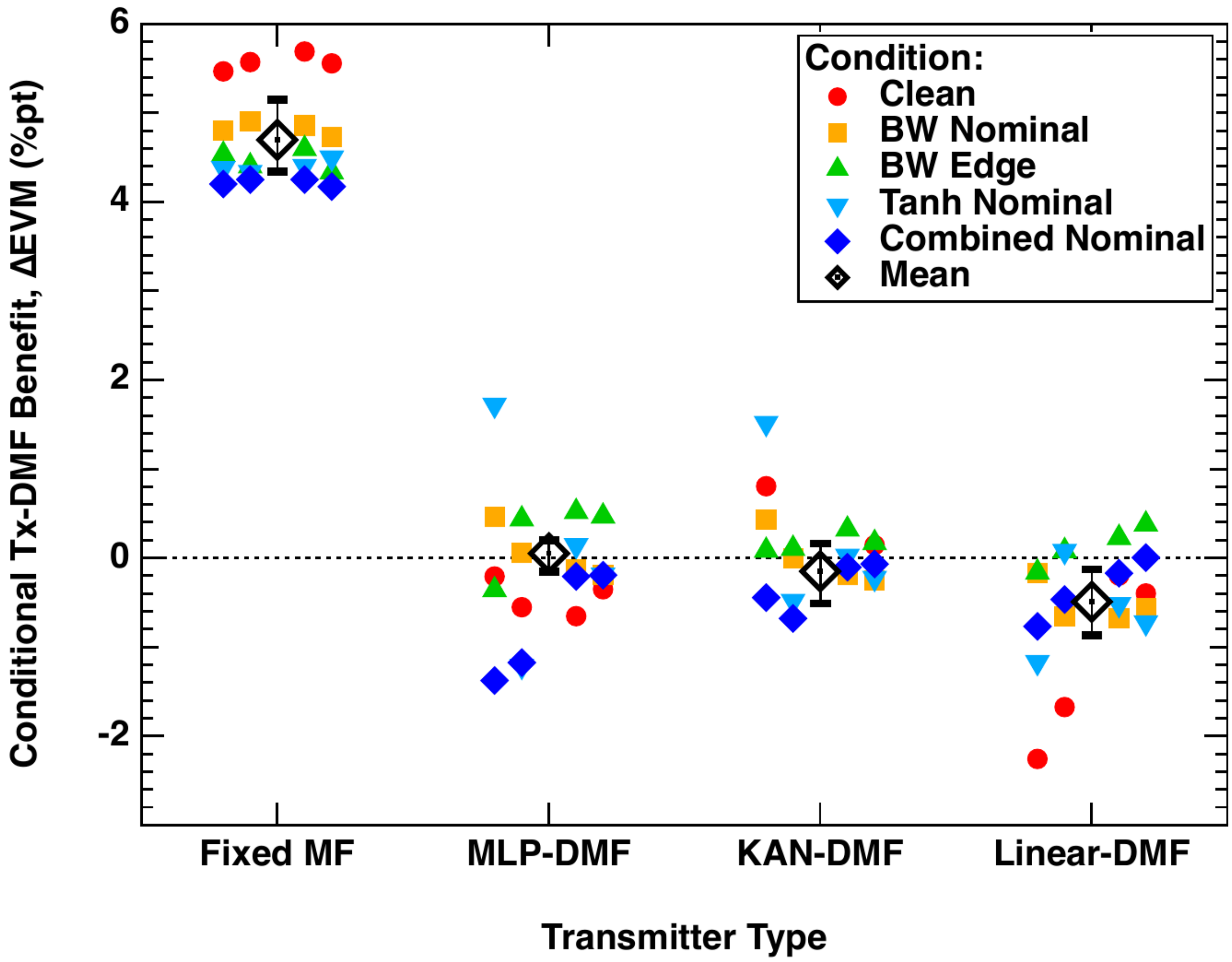

**Fig. S6. Moderate-condition factorial interaction.** Conditional transmitter benefit in EVM percentage points for a fixed matched-filter receiver and KAN-, MLP-, or linear-controlled Rx-DMF. Each colored point is one of the five prespecified physical condition clusters; black diamonds show their means. The fixed Tx-DMF pulse provides a consistent benefit with the fixed receiver, whereas its additional contribution is centered close to zero after nonlinear receiver adaptation and is mildly negative with linear control under these moderate conditions.

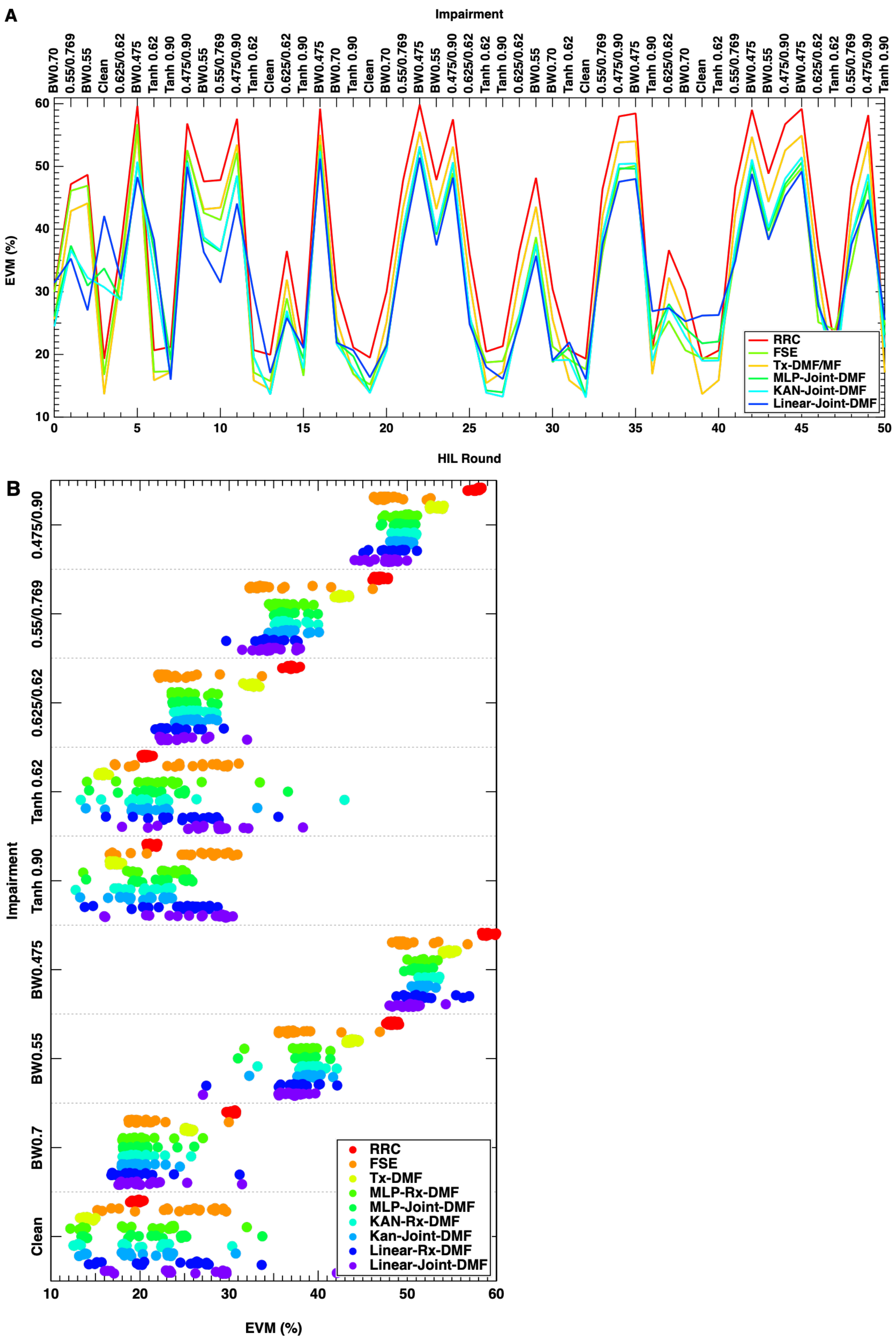

**Fig. S7. Adversarial rapid-switching stress test.** (A) EVM during the first 50 rounds of a 180-round sequence in which impairment condition changes nearly every HIL round. Adaptive state persists without condition labels or transition resets. (B) Condition-resolved EVM across the complete sequence for the fixed matched filter, FSE, transmitter-only DMF, and receiver-only and two-sided KAN-, MLP-, and linear-controlled systems. The experiment tests bounded state carry-over under hostile switching; it is not interpreted as a convergence experiment or a model of realistic channel evolution. Combined-condition labels x/y denote bandwidth fraction x followed by nonlinear strength y

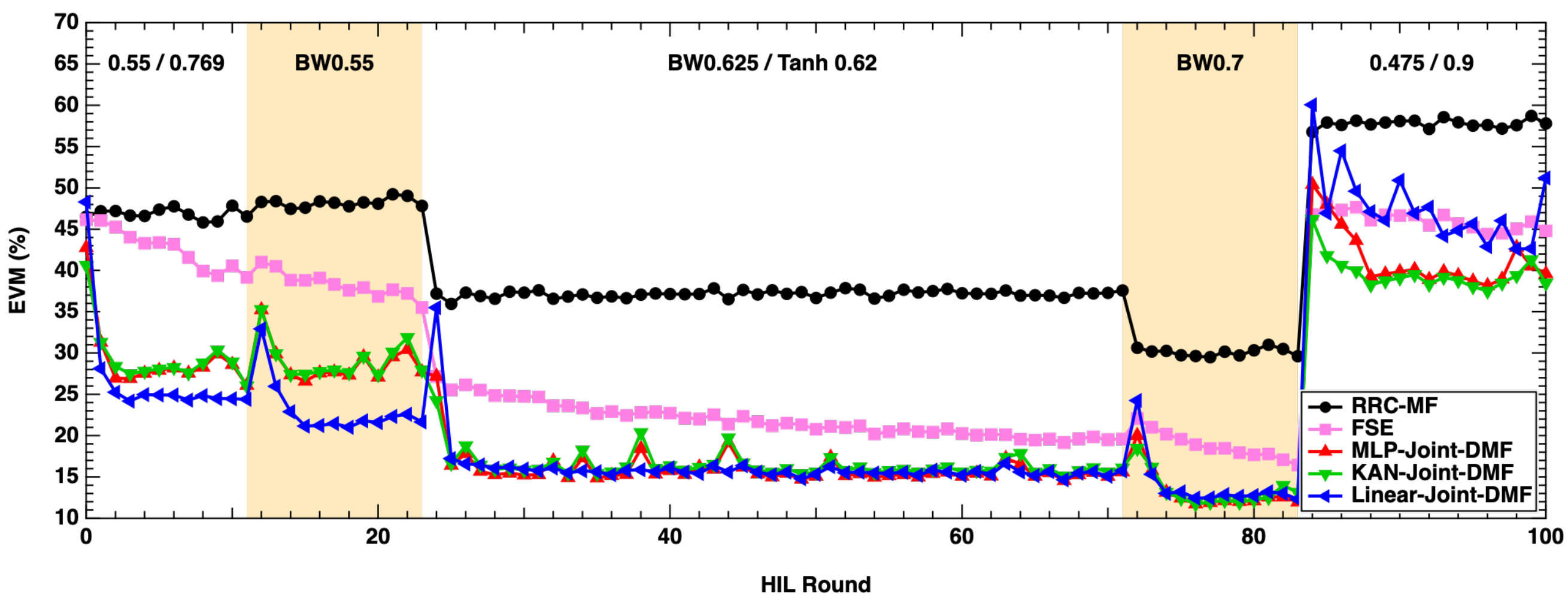

**Fig. S8. Two-sided trajectories during randomised stationary dwells.** Complete EVM trajectories of the fixed matched-filter baseline, persistent FSE, and two-sided KAN-, MLP-, and linear-controlled systems during the 108-round randomised dwell experiment. The five physical conditions remain fixed for 12, 12, 48, 12, and 24 consecutive rounds, respectively. No adaptive state is reset at a dwell boundary. Combined-condition labels x/y denote bandwidth fraction x followed by nonlinear strength y.

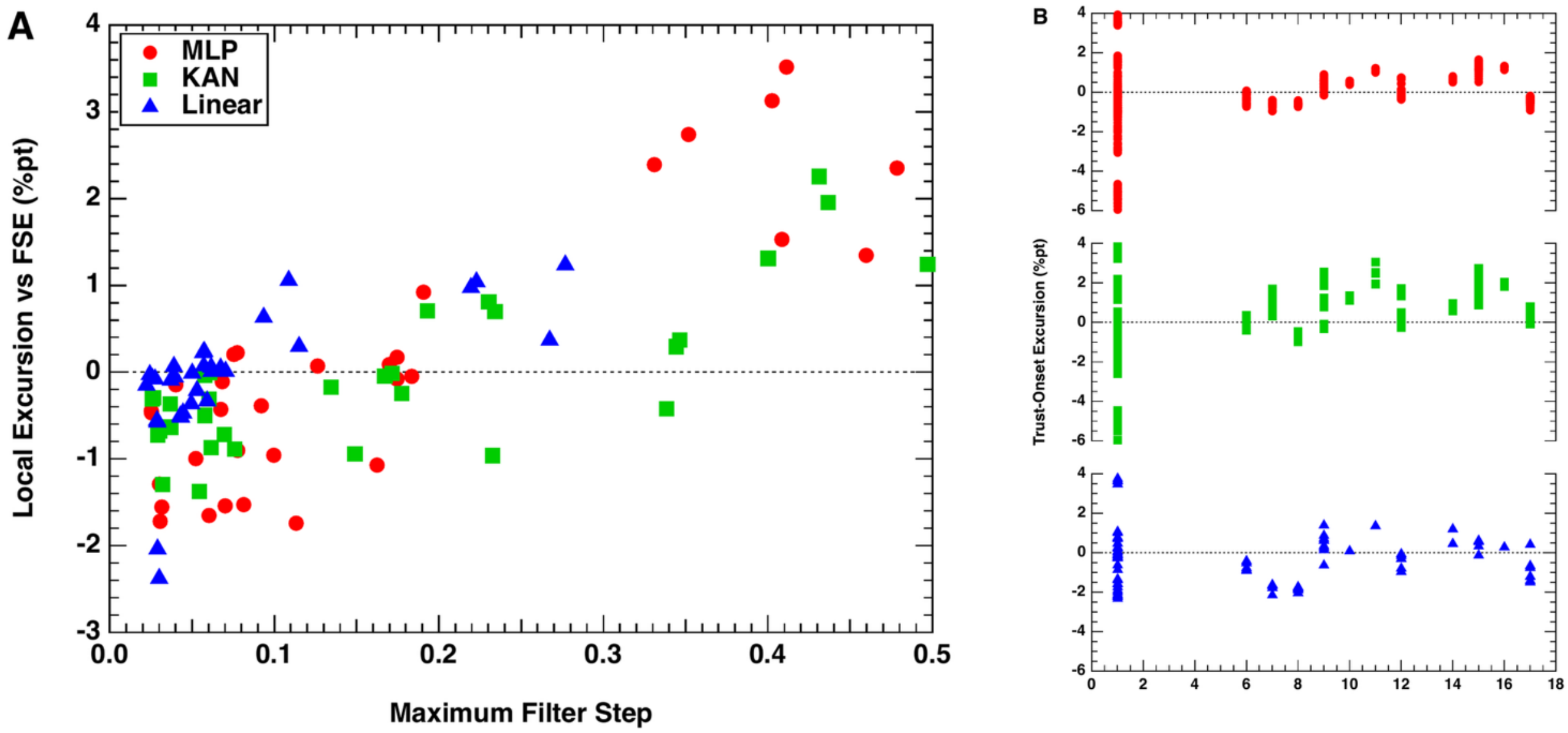

**Fig. S9. Adaptation-state origin of reproducible transient excursions.** (A) Local block-level DMF-minus-FSE excursion versus the maximum filter-step norm in the two clean-condition C12 physical repeats. (B) Excursion at the first block for which the prespecified 24-block grace interval has expired and the late-trust constraint becomes active (grace age 25), shown separately for MLP, KAN, and linear control. The local excursion is the current DMF-minus-FSE residual minus the mean of its immediately preceding and following residuals within the same persistent stream. Points include repeated seeds, captures, roll-offs, and blocks and are therefore descriptive rather than independent experimental replicates. The alignment supports attribution to structured adaptation dynamics but is not a no-update counterfactual test.

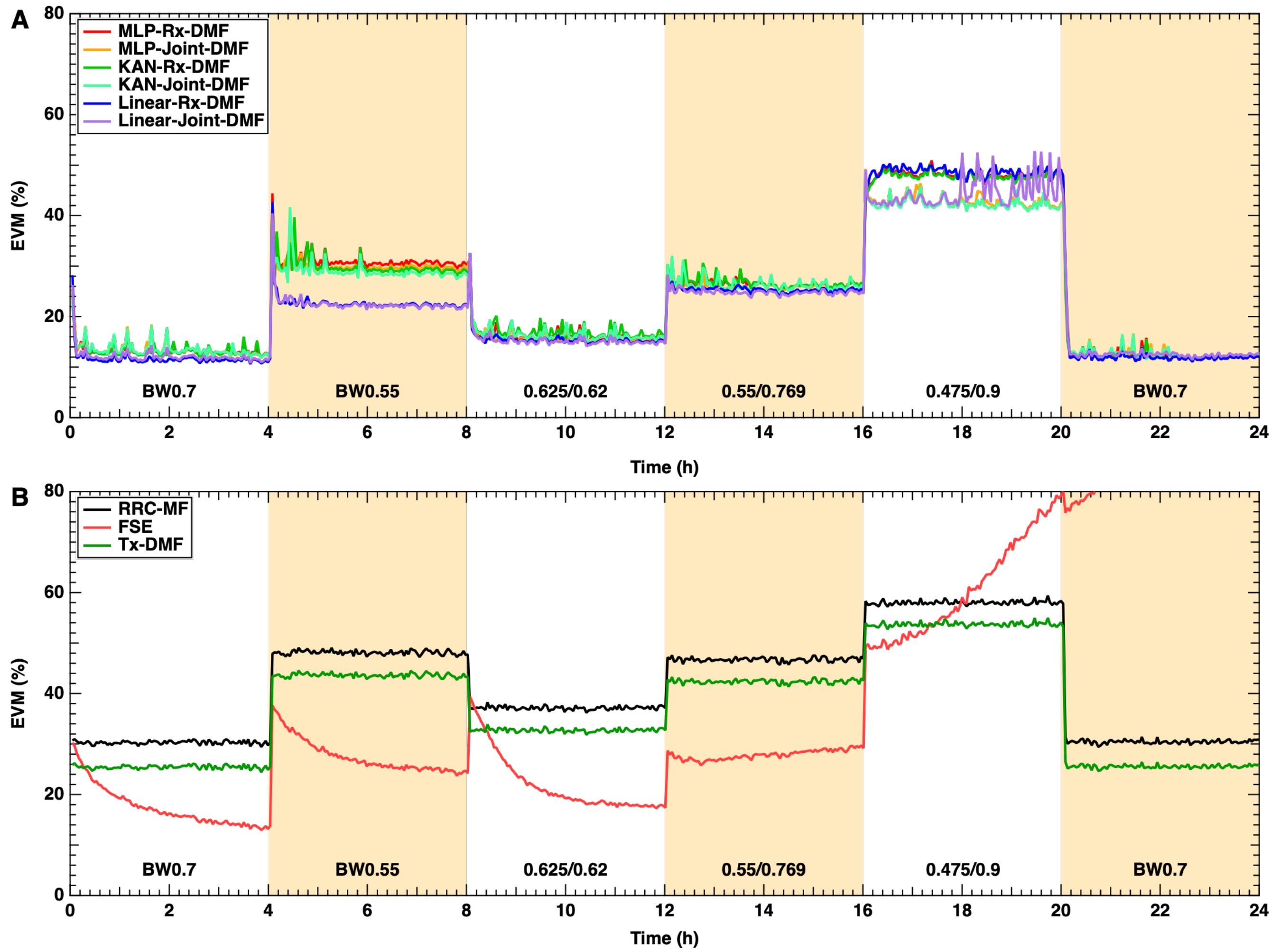

**Fig. S10. Complete receiver-only, two-sided, and fixed-filter trajectories over 24 hours.** (A) KAN-, MLP-, and linear-controlled Rx-DMF and joint Tx-DMF/Rx-DMF trajectories across all six physical channel epochs. (B) Fixed matched-filter, persistent FSE, and transmitter-only Tx-DMF/MF trajectories from the same uninterrupted run. Each controller trace is the per-round mean across four persistent seeds; all 542 completed rounds are shown. Combined-condition labels x/y denote bandwidth fraction x followed by nonlinear strength y.

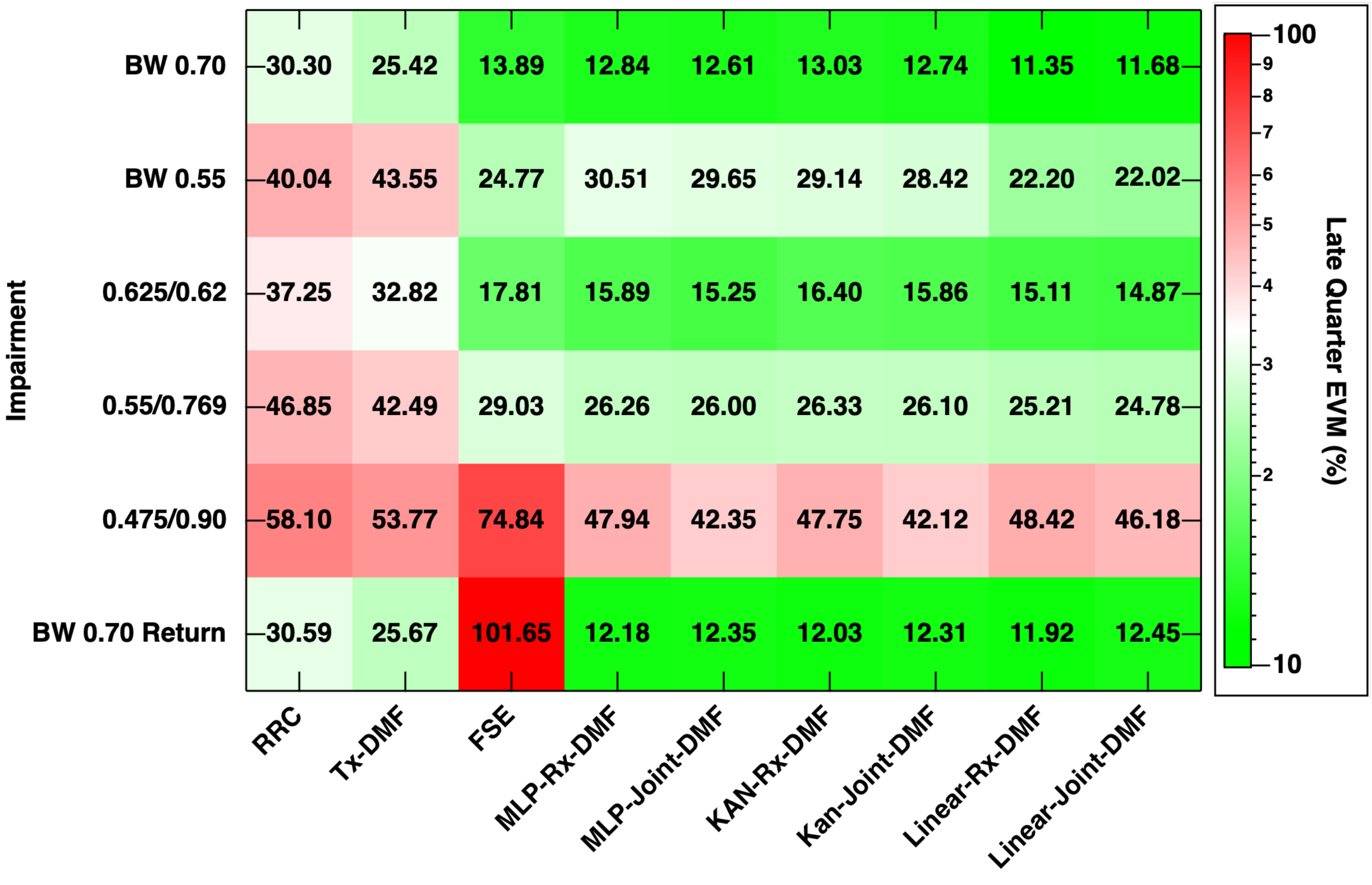

**Fig. S11. Condition-resolved late-quarter performance during 24-hour operation.** Late-quarter EVM for every fixed-filter, receiver-only, and two-sided system in each of the six physical channel epochs. Each value is the arithmetic mean over the prespecified final quarter of that epoch. The repeated bandwidth-0.70 visit provides the long-horizon return control and exposes persistent FSE divergence after severe intervening distortion. Combined-condition labels x/y denote bandwidth fraction x followed by nonlinear strength y.

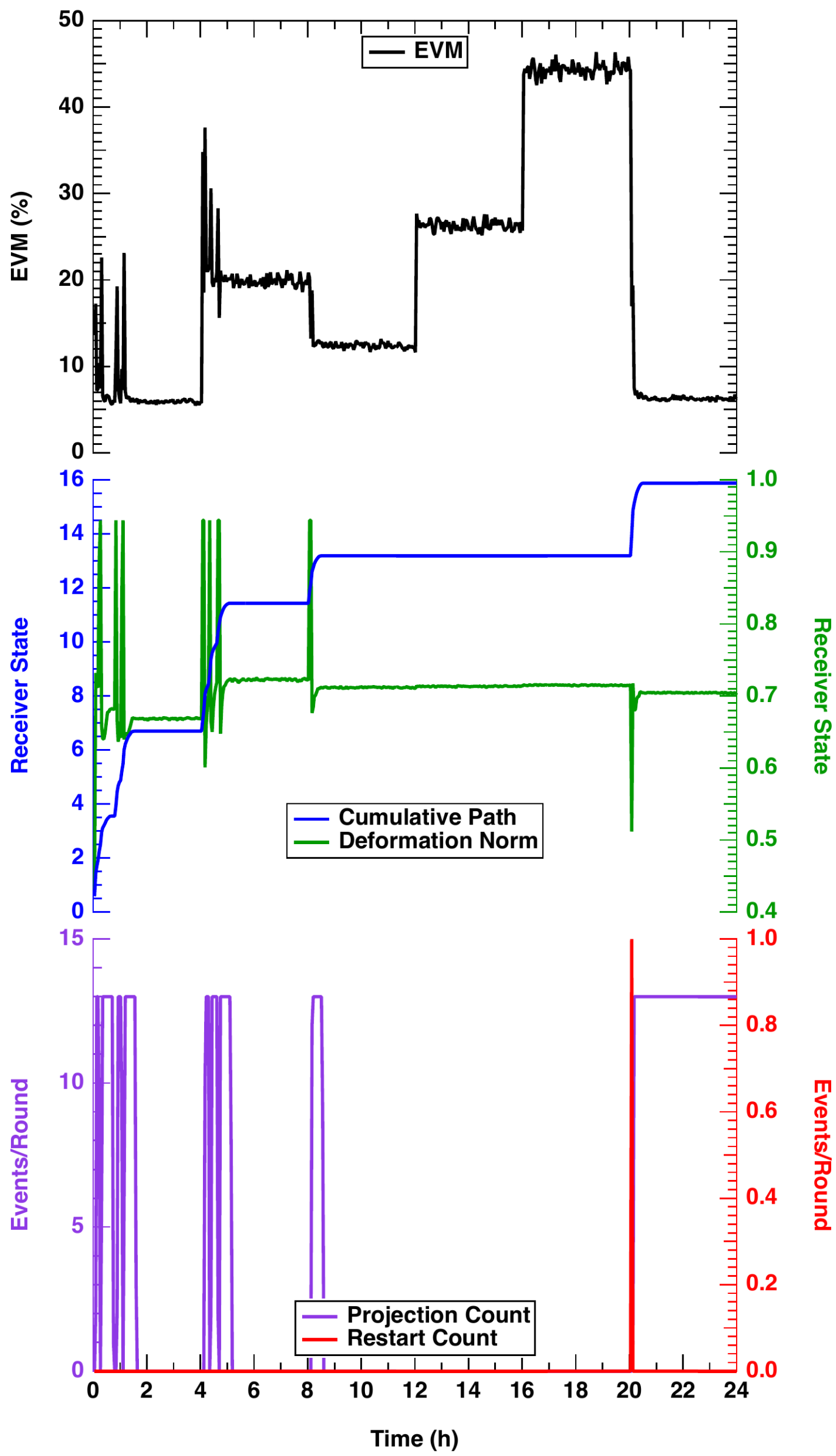

**Fig. S12. Representative persistent receiver-state dynamics.** Block EVM, cumulative filter-path length, deformation norm, late-trust projection count, and restart count for one representative RRC-transmitter PR4 combined link controlled by the MLP (seed 3234). The trajectory shows continued movement within a bounded deformation neighborhood across the six physical epochs. Projection and restart counts are internal stabilisation events, not scheduled responses to condition transitions.